\documentstyle[aps,12pt,epsf,manuscript]{revtex}
\begin{document}

\draft
\tightenlines
\preprint{KUNS 1605}
\title{	Nonequilibrium Weak Processes in Kaon Condensation II \\ 
--- Kinetics of condensation ---}
\author{Takumi Muto\thanks{p21mutot@cc.it-chiba.ac.jp}}	
\address{Department of Physics, Chiba Institute of Technology, 
2-1-1 Shibazono, Narashino, Chiba 275-0023, Japan}
\author{Toshitaka Tatsumi\thanks{tatsumi@ruby.scphys.kyoto-u.ac.jp}}   
\address{Department of Physics, Kyoto University, Kyoto 606-8502, Japan }
\author{Naoki Iwamoto\thanks{iwamoto@uoft02.utoledo.edu} }
\address{Department of Physics and Astronomy,
 The University of Toledo, Toledo, Ohio 43606-3390, U.S.A.}

\maketitle

\begin{abstract}
The kinetics of negatively charged kaon 
condensation in the early 
stages of a newly born neutron star is considered. 
The thermal kaon process, in which kaons are thermally produced by
nucleon-nucleon collisions, is found to be dominant throughout the
equilibration process. Temporal changes of the order parameter of the
condensate and the number densities of the chemical species are obtained 
from the rate equations,  which include the thermal kaon reactions 
as well as the kaon-induced Urca  and the modified Urca  reactions. 
It is shown that the dynamical evolution of the condensate is characterized 
by three stages: the first, prior to establishment of a condensate, 
the second, during the growth and subsequent saturation of the condensate, 
 and the third, near chemical equilibrium. 
The connection between the existence of a soft kaon mode 
and the instability of the noncondensed state is discussed. 
Implications of the nonequilibrium process on the possible 
delayed collapse of a protoneutron star are also mentioned. 
\end{abstract}

\newpage
\section{Introduction}
\label{sec:intro}

The possibility that baryonic matter may undergo a  
transition to a new hadronic phase consisting of a 
condensate of negatively charged kaons ($K^-$) 
in high density matter has been extensively discussed
\cite{kn86,t95,fmmt96,l96,pbpelk97,blr98}.  The kaon condensation may be 
formulated on the basis of 
 SU(3)$_{\rm L} \times$ SU(3)$_{\rm R}$ current algebra and PCAC 
in a model-independent way\cite{t95,fmmt96}. 
Within this framework, the $s$-wave kaon-condensed state
 $|\theta\rangle$ is generated by a chiral rotation of 
the normal state $|{\rm normal}\rangle$ as $ |\theta\rangle=
\hat U_K |{\rm normal}\rangle$ , where $\displaystyle
\hat U_K=\exp(i\mu_K t\hat Q)\exp(i\theta\hat Q_4^5) $
is a unitary operator. 
Here $\mu_K$ is the kaon chemical potential, 
$\theta$ the chiral angle, which is the order parameter 
of the condensate, and $\hat Q$ ($\hat Q_4^5$) is 
the electromagnetic charge operator (the axial charge operator).   
The classical $K^-$ field is then written as 
$\langle K^-\rangle\equiv\langle\theta |K^-|\theta \rangle =(f/\sqrt 2)
\sin\theta\exp(-i\mu_K t)$ 
with $f$(=93 MeV) the meson decay constant.

Kaon condensation is a form of Bose-Einstein condensation (BEC): 
 the lowest excitation energy of the $K^-$ decreases in a nuclear medium 
with an increase in the baryon number density 
due to an attractive $s$-wave
kaon-nucleon  ($KN$) interaction. When the lowest excitation energy 
becomes equal to the charge chemical potential, the distribution function
for the $K^-$ diverges and a macroscopic number  of kaons appear.  
The relevant $s$-wave $KN$ interaction  comes mainly  from the scalar interaction
(the $KN$ sigma term, $\Sigma_{\rm KN}$)  and the vector interaction 
(the Tomozawa-Weinberg term)\cite{t95,fmmt96,l96}. 

 The baryon number density for the onset of kaon condensation 
in stable neutron-star matter has been predicted to be 3$-$4 times  
the nuclear matter density $n_0$ (=0.16 fm$^{-3}$), 
depending on the magnitude of $\Sigma_{\rm KN}$.  
Motivated by studies of kaon condensation, the in-medium 
$KN$ interaction  has recently been elaborated both theoretically 
and experimentally, through $KN$ scattering\cite{k94,ww97,l98}, 
kaonic atoms\cite{bfg97}, and
kaon-antikaon production via relativistic 
nucleus-nucleus collisions\cite{sbmb97,llb97,cb99,ss99,slk99}. 
The experimental results suggest a substantial decrease 
in the antikaon effective mass, which in turn favors the possible 
existence of a kaon condensate in the core of a neutron star. 

The appearance of the kaon-condensed phase in a neutron star 
would lead to softening of the equation of state (EOS)\cite{fmmt96,llb97,tpl94}
and  also accelerate cooling of neutron stars\cite{tpl94,bkpp88,t88,pb90,fmtt94}. 
The softening of the EOS associated with kaon condensation has been 
used to construct a model in which the collapse of a hot protoneutron star 
to a black hole is delayed\cite{bb94,bst96,h99}
\footnote{Similar delayed collapse
mechanisms of a hot protoneutron star  to a more compact
neutron star or to a black hole, attributed to other hadronic phase
transitions,  have been considered by some
authors\cite{pbpelk97,tt88,ts88,bh89,kj95,p99}. };  
initially, there  is no kaon
condensation in protoneutron stars, since the neutrino degeneracy and thermal effects
raise the threshold density for kaon  condensation. A kaon condensate is formed
during the evolution 
 into a compact neutron star through deleptonization and cooling. 
 The maximum mass of the kaon-condensed neutron star 
is estimated to be $\sim $1.5 $M_\odot$ with $M_\odot$ the solar mass
\cite{fmmt96,llb97,tpl94}, which is smaller than the maximum mass 
of ordinary neutron stars ($\sim 2.0 M_\odot$)\cite{apr98} 
because of the significant softening of the EOS
 due to the appearance of the condensates. 
If the mass of a protoneutron star after deleptonization or cooling exceeds 
the maximum mass of the kaon-condensed neutron star, the star 
eventually collapses to a black hole.  
Such  a scenario for the formation of a low-mass
black hole (1.5$-$2.0 $M_\odot$) has been presented by Brown and
Bethe\cite{bb94} to interpret the absence of a pulsar signature 
in SN 1987A\cite{p95}.  
Following their scenario,  Baumgarte et al. made a numerical simulation 
for the delayed collapse of a protoneutron star  by the use of an  EOS including the
phase transition to the  kaon-condensed phase\cite{bst96}. 

In these studies, the equilibrated EOS 
with kaon condensation was used. The kaons are
created through  {\it weak interactions} that change strangeness during the
appearance and the growth of the condensate, 
and the time scale for the weak
reactions are much larger than those for the strong and electromagnetic 
reactions which are responsible for the thermal equilibration of the system. 
In addition,  there are at least three time scales which characterize the dynamical
evolution  of a newly born neutron star: the time scale for gravitational collapse 
(of order of a millisecond)  and those  
for deleptonization and initial cooling 
(of order of a second and ten seconds, respectively).  
If the weak reactions which are responsible for the formation 
of a condensate proceed on a longer time scale than these time scales, 
they may control the dynamical evolution of a neutron star  
just after a supernova explosion. In this case, the 
nonequilibrium processes in kaon condensation which are brought about 
by the weak reactions must be considered.  

The time evolution of the system is determined by the Hamiltonian, 
\begin{equation}
\hat H=\hat H_{\rm strong}+\hat H_{\rm weak} \ , 
\label{hamiltonian}
\end{equation}
with the strong (weak) Hamiltonian 
$\hat H_{\rm  strong}$ ($\hat H_{\rm weak}$). In our case the state
$|\theta(t),t\rangle$ is specified by the parameters, 
$\theta$, $\mu_n$ ($\mu_p$) the neutron chemical potential (the proton 
chemical potential), $\mu_K$, and $\mu_e$ the electron chemical potential, 
for given temperature $T$ and baryon number density $n_B$. Alternatively 
we may choose another set of operators, $\hat\theta$, 
the number densities $\hat n_i$
($i=p,n,K,e$) to specify the state. If the weak Hamiltonian 
$\hat H_{\rm weak}$ is negligibly small, then the time scale of the weak
interaction is  very large ($\tau_{\rm weak}\gg \tau_{\rm strong}$). 
In such a case, it is obvious that
all the number densities are time-independent, $\dot{\hat n}_i$=0, 
which in turn implies $\dot{\hat \theta}$=0, or they can commute with
$\hat H_{\rm strong}$, $[\hat n_i,\hat H_{\rm strong}]$=0, due to baryon
number, strangeness, charge and lepton number conservations. 
Thus we can find the simultaneous eigenstate of 
$\hat H_{\rm strong}$ and $\hat n_i$. Hence, if we apply the
adiabatic (Born-Oppenheimer) approximation, 
then the time-dependence of 
$n_i(t)\equiv \langle\theta(t), t|\hat n_i|\theta(t), t\rangle$ 
is determined by the weak interaction Hamiltonian $\hat H_{\rm weak}$, 
which leads to the rate equations. With $n_i(t)$ taken to be a function of time, 
it is possible to fix the time and 
construct an instantaneous eigenstate of
$\hat H_{\rm strong}$ and $\hat\theta$, by considering the
extremum conditions for the thermodynamic potential $\Omega$ 
given only by the strong interaction Hamiltonian $\hat H_{\rm strong}$, 
\begin{equation}
|\theta(t),t\rangle\simeq |\theta(t), n_K(t), n_n(t), n_p(t), n_e(t)\rangle . 
\label{state}
\end{equation}
Solving the rate equations then allows the temporal change of $n_i$ 
to be determined, while the time dependence of the order parameter $\theta$ is
determined by use of the  thermodynamic potential $\Omega$. 

The nonequilibrium processes in the kaon-condensed state 
$|\theta(t), t\rangle$ have been previously investigated in \cite{mti97}, 
with account taken of both the kaon-induced Urca process 
and the modified Urca process. The former process 
(abbreviated to KU-F and KU-B,
where  F and B stand for forward and backward reactions, respectively) is 
represented by the reactions 
\begin{mathletters}\label{ku}
\begin{eqnarray}
& & n(p)\rightarrow n(p) +e^- +\bar\nu_e, \label{fku} \\
& & n(p)+e^- \rightarrow n(p)+\nu_e \ , \label{bku}
\end{eqnarray}
\end{mathletters}
and provides the most 
efficient cooling mechanism for the star via neutrino and 
antineutrino emissions\cite{bkpp88,t88}. 
In terms of the condensate
$\langle K^-\rangle$,  by which the system is supplied with energy and the reactions
become kinematically possible, 
the KU process (\ref{ku}) can be written  symbolically as 
$n(p)+\langle K^-\rangle\rightarrow n(p)+e^- +\bar\nu_e$, 
$ n(p)+e^- \rightarrow n(p)+\langle K^-\rangle+\nu_e $. 
The modified Urca process (MU-F and MU-B)\cite{fm79,m87}, represented by 
the reactions 
\begin{mathletters}\label{mu}
\begin{eqnarray}
& & n+n\rightarrow n+p +e^- +\bar\nu_e,\label{fmu} \\
& & n+p+e^- \rightarrow n+n+\nu_e \ , \label{bmu}
\end{eqnarray}
\end{mathletters}
is a standard cooling process for a normal neutron star.\footnote{
Throughout this paper, we do not take into account 
the direct Urca process in the normal phase\cite{b81,lpph91} or in the
kaon-condensed phase\cite{tpl94,fmtt94},
 and also omit other weak reactions associated
with hyperons\cite{pplp92},  taking the same standpoint as in Paper I. }
Because the transition matrix element for KU is  proportional to $\sin\theta$, 
this process is operative  only in the presence of a condensate. 
Thus, spontaneous creation of a condensate from the noncondensed state 
cannot occur solely by way of the KU and MU reactions, and, for this reason 
in Ref.\cite{mti97}, 
 a small seed of the condensate ($\theta\neq 0$) was put in by hand. 

In the noncondensed state, kaons are produced thermally by the
weak reactions
\begin{mathletters} \label{kt}
\begin{eqnarray}
& & n+n \rightarrow  n+p+K^- , \label{fkt} \\
 & &n+p+K^- \rightarrow n+n  \ ,    \label{bkt} 
\end{eqnarray}
\end{mathletters} 
where  a spectator neutron must take part in the reactions such that 
 the kinematical condition in a degenerate Fermi system is satisfied. 
The thermal kaons are then converted to a condensate. 
Thus we can discuss the onset and the growth of a kaon condensate 
consistently without any ad hoc seed (see Sec.\ref{sec:kinetics}), 
in which case the process (\ref{kt}) is primarily responsible 
for the onset of condensation.
We will refer to the reaction (\ref{kt}) as the thermal kaon process, and 
abbreviate them to KT-F and KT-B, respectively. 

In a previous paper \cite{mti99-1} (Paper I), 
we have obtained the reaction rate for KT, and discussed the effects  
 of thermal kaons on the KT reaction rate in both the noncondensed and 
condensed states. We have also compared the KT reaction rate 
with the rates for the KU and MU reactions and obtained the following 
results:  (i) In the noncondensed state,  where the system is far from
chemical equilibrium,  hard thermal  kaons with large momenta 
make the major contribution to the KT reaction rate, whereas  
in the condensed state, the soft thermal kaons, which appear 
as a Goldstone mode from the spontaneous breaking of $V$-spin symmetry,  
contribute significantly to the KT reaction rate. 
(ii) The KT reaction rate is larger than 
the rates for the KU and MU reactions over the relevant temperatures 
and baryon number densities. (iii) 
The KT process is dominant in the kaon-condensed state in chemical
equilibrium as well as in the noncondensed state, and may determine the
evolution of the system. 

Based on the above results for the KT reactions, we in this paper, 
investigate the nonequilibrium weak processes in kaon condensation 
by taking into account the KT process as well as the KU and MU processes, 
and clarify the effects of thermal kaons on the kinetics of condensation. 
Assuming that the system is in thermal equilibrium, we solve the
rate  equations given by the relevant weak reactions 
(\ref{ku})$-$(\ref{kt}), and discuss the temporal behavior 
of the number densities of the chemical species, the order parameter of the
condensate,  and other physical quantities.  In general, the energy, produced 
in the course of the nonequilibrium process, is dissipated into 
the surroundings, resulting in a rise in the temperature of the system. 
For simplicity, however, the temperature is kept constant  
throughout this paper. 
We find two characteristic time scales for the onset of a  condensate 
and its subsequent growth.  The magnitudes of these time scales are 
compared with other time scales characterizing the collapse of a 
newly-born  neutron star,  and possible effects on stellar collapse
 are discussed. 

The paper is organized as follows: 
The description of the nonequilibrium state based on the thermodynamic
potential is addressed in Sec. \ref{sec:description}, 
and the formulation for obtaining the rate equations is given in
Sec. \ref{sec:kinetics}. The numerical 
results are then presented in Sec. \ref{sec:result}, and 
summary and concluding remarks are given in Sec. \ref{sec:summary}. 
In Appendix A, an asymptotic behavior of the system 
near chemical equilibrium is discussed. Specifically, 
an expression for the relaxation
time  near chemical equilibrium is derived analytically. 

\section{Description of the Nonequilibrium State }
\label{sec:description}
\subsection{Thermodynamic potential}
\label{subsec:thermo}

Here we define the physical conditions for a nonequilibrium 
kaon-condensed state, $|\theta(t), t\rangle$.\footnote{
We use the units in which $\hbar$=$c$=$k_B$=1 throughout this paper.}
The system is described by a thermodynamic potential $\Omega$. 
The microscopic quantities, such as the excitation energies of the 
baryons and the thermal and condensed kaons, are  
determined by the strong and electromagnetic 
interactions among the kaons, baryons and leptons. 
The time scales for these interactions are much smaller 
than those for weak reactions. 
Therefore, in setting up the kinetic equations, that determine 
the relatively slow change of chemical composition via weak reactions,  
the system may be assumed  to be in thermal equilibrium, 
maintained by much faster strong and electromagnetic reactions. 
As a consequence, the physical quantities for the  nonequilibrium state
evolve adiabatically,  adjusting to the gradual change 
of the chemical composition  brought about by the weak processes. 

Assuming thermal equilibrium, 
we adopt the thermodynamic potential 
$\Omega$ of the kaon-condensed phase which was derived 
by Tatsumi and Yasuhira from chiral symmetry\cite{ty98}, including the 
thermal and quantum fluctuations around a condensate. 
In the present work, we shall neglect, for the sake of simplicity,  
both zero-point and thermal fluctuations in the thermodynamic variables 
except for the thermal contribution to the kaon number density, which  
allows us to see the effects of thermal kaons on the kinetics of condensation. 
Thus our expression for $\Omega$ corresponds to the heavy-baryon limit 
for the nucleons\cite{ty98}. With these assumptions, the thermodynamic
potential per unit volume
$\Omega/V$  of the kaon-condensed phase is written as    
\begin{equation}
\Omega/V=(\Omega_C+\Omega_K+\Omega_N+\Omega_e)/V \ , 
\label{omega}
\end{equation}
where $\Omega_C$ and $\Omega_K$ are the potentials arising from 
a condensate and thermal fluctuations for the kaons, respectively,  and 
$\Omega_N$ and $\Omega_e$ are the corresponding potentials  
for the nucleons and the electrons, respectively.  
Here, $\Omega$, the number density
$n_i$ and the chemical potential $\mu_i$ ($i=p,n,e^-,K$) 
are related to each other by $n_i=-\partial(\Omega/V)/\partial \mu_i$. 
Specifically, the classical contribution from the condensate, $\Omega_C$, is 
given by 
\begin{equation}
\Omega_C/V=f^2m_K^2 (1-\cos\theta)
-\frac{1}{2}f^2\mu_K^2\sin^2\theta \ , 
\label{omegac}
\end{equation}
where $m_K$ is the free kaon mass (= 494 MeV). 
In contrast, the thermal kaon contribution, $\Omega_K$, is given by  
\begin{equation}
\Omega_K/V=T\int\frac{d^3
p_K}{(2\pi)^3}\ln\Big(1-e^{-\lbrack\omega_+({\bf p}_K)
+\mu_K\rbrack/T}\Big)\Big(1-e^{-\lbrack\omega_-({\bf p}_K)
-\mu_K\rbrack/T}\Big) \ , 
\label{omegak}
\end{equation}
with $\omega_+({\bf p}_K)$ [$\omega_-({\bf p}_K)$] 
the excitation energy for $K^+$ ($K^-$) 
(cf. Sec. \ref{subsec:thermal} for the explicit expressions 
for $\omega_\pm ({\bf p}_K)$ ), and the nucleon contribution 
$\Omega_N$ is written in the standard form:
\begin{equation}
\Omega_N/V={\cal E}_N-\mu_p n_p-\mu_nn_n 
\label{omegan}
\end{equation}
where the energy density for the nucleons is given by 
\begin{equation}
{\cal E}_N={3\over
5}\frac{(3\pi^2)^{2/3}}{2m_N}\Big(n_p^{5/3}+n_n^{5/3}\Big)
-f^2(\sigma+2b\mu_K)(1-\cos\theta) +V_{\rm
sym}(n_B)\Big(n_p-n_n\Big)^2/n_B \ . 
\label{en}
\end{equation}
The first term in the r.h.s. of Eq. (\ref{en}) is the nucleon  
Fermi-gas energy density, with $m_N$ the nucleon mass, and 
the quantity $\sigma$ in the second term defined as 
$\sigma\equiv n_B\Sigma_{\rm KN}/f^2$, 
and $\displaystyle b\equiv\Big(n_p+\frac{1}{2}n_n\Big)/(2f^2)$ is 
the $V$-spin density [see Sec. \ref{subsec:times} for specific values]. 
The terms involving $\sigma$ and
$b$, respectively,  come from the $s$-wave $K^-N$ interaction 
given by the $KN$ sigma term and the Tomozawa-Weinberg term. 
The expression for the energy contribution 
from kaon-kaon and  kaon-nucleon interactions 
is essentially determined in a model-independent way by chiral symmetry. 
The third term in Eq.(\ref{en}) is the potential energy 
contribution to the symmetry energy.   
For $V_{\rm sym}(n_B)$, we use the expression given by Prakash et al.
\cite{pal88}:
\begin{equation}
V_{\rm sym}(n_B)=\Big\lbrack
S_0-(2^{2/3}-1)\frac{3}{5}\epsilon_{F,0}\Big\rbrack F(n_B) \ ,
\label{symmetry}
\end{equation}
where $S_0$ (= 30 MeV) is the empirical symmetry energy, and 
$\epsilon_{F,0}$ is the Fermi energy in the symmetric nuclear matter 
at the density $n_0$, with the function $F(n_B)$ taken as 
$F(n_B)=n_B/n_0$ for simplicity.  
The chemical potentials for the proton and the neutron, respectively, in
(\ref{omegan}) are given by $\mu_p=\partial{\cal E}_N/\partial n_p$ 
and $\mu_n=\partial{\cal E}_N/\partial n_n$, respectively, so that 
the difference between
$\mu_p$ and $\mu_n$  in the condensed state is written as 
\begin{eqnarray}
\mu_p-\mu_n&=&\frac{(3\pi^2 n_p)^{2/3}}{2m_N}-
\frac{(3\pi^2n_n)^{2/3}}{2m_N}+4V_{\rm sym}(n_B)\cdot
(n_p-n_n)/n_B \cr
&-&\frac{1}{2}\mu_K(1-\cos\theta)  \ . 
\label{ceppn}
\end{eqnarray}
For $\Omega_e$, we use the ultrarelativistic form for electrons, 
\begin{equation}
\Omega_e/V=\frac{\mu_e^4}{4\pi^2}-\mu_e n_e
=-\frac{\mu_e^4}{12\pi^2} \ , 
\label{omegae}
\end{equation}
where $\mu_e$ is the electron chemical potential and
$n_e$ [=$\mu_e^3/(3\pi^2)$] is the electron number density. 
Note that $\mu_e\neq \mu_K$ when the system is 
not in chemical equilibrium. 

\subsection{Thermal kaon excitation }
\label{subsec:thermal}

As can be seen in Eq.(\ref{omega}), only the thermal kaon excitations  
constitute the thermal fluctuations 
in the thermodynamic potential $\Omega$,  
because they are responsible for 
the formation of a condensate through the weak reactions 
KT[(\ref{kt})] . The excitation energy for the thermal kaons, 
$\omega_\pm({\bf p}_K)$,  appearing in $\Omega_K$ [Eq.(\ref{omegak}) ], 
is obtained from the dispersion relations for kaon modes 
in the condensed state. The expression for
$\omega_\pm({\bf p}_K)$ depends on the method for treating the fluctuations 
around the condensate\cite{ty98,te97}, but numerical results 
have shown that there is little difference between the two 
methods used in Refs.\cite{ty98} and \cite{te97}. 
 Later on we use the result of Ref.\cite{ty98}:
\begin{equation}
\omega_{\pm}({\bf p}_K)\simeq\pm\Big\lbrace
b+\mu_K(\cos\theta-1)\Big\rbrace +\Big\lbrack {\bf
p}_K^2+(b^2+\widetilde{ m_K^{\ast 2}})
\Big\rbrack^{1/2} \ ,  
\label{cenergy}
\end{equation}
where $\widetilde {m_K^{\ast 2}}\equiv m_K^{\ast 2}\cos\theta
=(m_K^2-\sigma)\cos\theta$, and $m_K^{\ast}$ 
denotes the effective kaon mass  
which is reduced due to  the $KN$ scalar interaction  
$\Sigma_{\rm KN}$\cite{fmmt96}. 
Using (\ref{cenergy}) one obtains the total kaon number density
from the relation, 
$n_K=-\partial(\Omega/V)/\partial \mu_K$: 
\begin{equation}
n_K=\zeta_K+n_K^T \ .
\label{nk}
\end{equation}
Here $\zeta_K$ is the condensed part of the kaon number density :
\begin{eqnarray}
\zeta_K&\equiv& -\partial(\Omega_C/V+\Omega_N/V)/\partial\mu_K \cr
&=&\mu_K f^2\sin^2\theta+2bf^2(1-\cos\theta)  \ ,
\label{nkc}
\end{eqnarray}
and $ n_K^T$ is the thermal part: 
\begin{eqnarray}
 n_K^T&=&-\partial(\Omega_K/V)/\partial \mu_K \cr
&=&{1\over (2\pi)^3}\cos\theta\int d^3p_K f_{K}({\bf p}_K)
\label{nkt}
\end{eqnarray}
with 
\begin{equation}
f_{K}({\bf p}_K)=
{1\over e^{[\omega_-({\bf p}_K) -\mu_K]/T}-1}
-{1\over e^{[\omega_+({\bf p}_K) +\mu_K]/T}-1}\ ,
\label{eqdistribution}
\end{equation}
where the first and second terms 
are the Bose-Einstein distribution functions for the $K^-$ 
and $K^+$ mesons, respectively. 
It is to be noted that the expression (\ref{nkt}) is slightly 
different from the usual one for the noncondensed  state 
by a reduction factor $\cos\theta$ due to the existence of the condensate. 
The expression (\ref{nkc}) is equivalent to the strangeness number density 
$\zeta_K=\langle K^-|{\hat S}|K^-\rangle$ with the strangeness operator 
${\hat S}\equiv 2({\hat Q}-{\hat I}_3-{\hat B})$.  

By the use of the classical field equation for
$\theta$, $\partial\Omega/\partial\theta=0$\cite{fmmt96,ty98}, 
\begin{equation}
\sin\theta\Big(\mu_K^2\cos\theta+2b\mu_K
-m_K^{\ast 2}\Big)=0 \ , 
\label{fieldeq}
\end{equation}
which is valid where fluctuations of kaons can be neglected\cite{ty98},  
it can be seen from (\ref{cenergy}) that, in the condensed phase 
($\theta\neq 0$), the lowest excitation energy of $K^-$ is equal to the kaon
chemical potential, i.e., 
$\omega_-({\bf p}_K=0)=\mu_K$. 
This soft mode results from the spontaneous $V$-spin symmetry breaking 
in the condensed phase (Goldstone mode), and leads to a  
divergence of $f_K({\bf p}_K)$ at ${\bf p}_K$=0 in the condensed state. 
On the other hand, in the limit $\theta=0$ in
Eq.(\ref{cenergy}), one obtains the excitation energy 
of kaons in the normal (noncondensed) phase: 
\begin{equation}
\omega_\pm({\bf p}_K)=\pm b+\Big\lbrack {\bf
p}_K^2+(b^2+ m_K^{\ast 2})
\Big\rbrack^{1/2} \ . 
\label{nenergy}
\end{equation}
In this case, there is a gap between the lowest excitation energy 
$\omega_-({\bf p}_K=0)$ and $\mu_K$,  so that  $f_K({\bf p}_K)$ has no
singularity. 
The appearance of a soft kaon mode is also related to an instability 
of the normal state with respect to the onset of a condensate. 
(See also Sec.\ref{subsec:times}.) 

\section{Kinetics of Condensation}
\label{sec:kinetics}

\subsection{Rate Equations}
\label{subsec:rate}

The thermally excited kaons, which are produced via the weak reactions KT, 
are converted into a condensate through kaon-nucleon and kaon-kaon 
scatterings: $KN\rightarrow\langle K\rangle N$, 
$KK\rightarrow K\langle K\rangle$.  
The time required for the conversion of the thermal kaons to a condensate 
through strong interaction collisions is negligible 
compared with the time scale governed by the weak reactions. This is
consistent with the assumption that the system is in
 {\it thermal equilibrium}, as discussed in Sec.\ref{subsec:thermo}. 
Thus, we put aside the detailed conversion mechanisms of the thermal kaons into a
condensate, and study the kinetics of condensation due to the weak reactions 
to obtain the typical time scales for the development of a condensate.  
With the kaon number density given by Eq.(\ref{nk}), 
one can see that the condensate appears 
spontaneously once the number density of the thermal kaons is saturated  
for a given temperature. With this simplification, we can describe 
the growth of the condensate 
during the whole nonequilibrium stage by following the kinetics 
of the condensate semiclassically, avoiding the discussion of 
quantum nucleation of the condensates.

In general, the heat released by 
 dissipation of energy in the system is expected to 
increase the temperature $T$.  The temporal change of the temperature,
$T(t)$ can be obtained from the equation that gives the rate of change of the
internal energy of the system: e.g., 
\begin{equation}
\partial E/\partial t=-\epsilon_\nu-\epsilon_{\bar\nu}
\label{dedt}
\end{equation}
with $\epsilon_\nu$, $\epsilon_{\bar\nu}$ the luminosities of neutrinos 
and antineutrinos, respectively, for the neutrino-free-streaming case. 
For simplicity, however, we
take the temperature to be constant during the nonequilibrium process. 
A more realistic calculation including the variation of the temperature 
will be discussed in the future. 

The number density of each chemical species is  determined  by the rate
equations, with the rates of change of the electron ($n_e$), the kaon ($n_K$), and
the proton ($n_p$) number densities given by 
\begin{mathletters}\label{rateeq}
\begin{eqnarray}
dn_e(t)/dt
&=&\Gamma^{\rm (KU-F)}(\xi^{({\rm KU})}(t),T)
-\Gamma^{\rm (KU-B)}(\xi^{({\rm KU})}(t),T) \cr
&+&\Gamma^{\rm (MU-F)}(\xi^{({\rm MU})}(t),T)
-\Gamma^{\rm (MU-B)}(\xi^{({\rm MU})}(t),T) \ ,
\label{erateeq} \\ 
dn_K(t)/ dt
&=&- \Gamma^{\rm (KU-F)}(\xi^{({\rm KU})}(t),T)
+\Gamma^{\rm (KU-B)}(\xi^{({\rm KU})}(t),T) \cr
&+&\Gamma^{\rm (KT-F)}(\xi^{({\rm KT})}(t),T)
-\Gamma^{\rm (KT-B)}(\xi^{({\rm KT})}(t),T) \ , 
\label{krateeq} \\
dn_p(t)/ dt
&=&\Gamma^{\rm (MU-F)}(\xi^{({\rm MU})}(t),T)
-\Gamma^{\rm (MU-B)}(\xi^{({\rm MU})}(t),T)  \cr
&+&\Gamma^{\rm (KT-F)}(\xi^{({\rm KT})}(t),T)
-\Gamma^{\rm (KT-B)}(\xi^{({\rm KT})}(t),T) \ , 
\label{prateeq}
\end{eqnarray}
\end{mathletters}  
where $\Gamma^{(\alpha)}$ are the reaction rates 
($\alpha$= KU, MU, KT), and `F' (`B') denotes the forward (backward)
process. As a consequence of baryon number conservation, 
the rate of change of the neutron number density
 is determined by $dn_p(t)/ dt$ through the equation 
$dn_n(t)/ dt=-dn_p(t)/ dt$. 

\subsection{Reaction rates}
\label{subsec:reaction}

Since the kaon-condensed state is obtained from the
underlying chiral symmetry
\cite{mti99-1}, the matrix elements for the relevant reactions 
can be calculated from the chirally-transformed weak Hamiltonian.     
The expressions for these reaction rates have been given in paper 
I \cite{mti99-1}.  Here we only show the results. 
For the forward KT process  (\ref{fkt}), denoted as KT-F, 
one obtains
\begin{mathletters}\label{frrfkt}
\begin{eqnarray}
\Gamma^{\rm (KT-F)}(\xi^{\rm (KT)},T)&=&
\frac{512}{9(2\pi)^7}\Bigg(g_AG_F{\widetilde f}^2f\sin\theta_C
\cos\theta_C\cos^3{\theta\over 2}
\frac{|{\bf p}_F(n)|^2}{|{\bf p}_F(n)|^2+m_\pi^2}\Bigg)^2 |{\bf p}_F(p) |\cr
&\times&(m_n^\ast)^3m_p^\ast T^5 
I^{({\rm KT})}(\xi^{\rm (KT)},T) \label{frrfkta} \\
&=&(4.0\times 10^{30})\bigg(\frac{|{\bf p}_F(p)|}{m_\pi}\bigg)
\bigg({m_n^\ast\over m_N}\bigg)^3
\bigg({m_p^\ast\over m_N}\bigg) 
\cos^6{\theta\over 2}T_9^5 I^{\rm (KT)}(\xi^{\rm (KT)},T) \cr 
& &\hspace{8.0cm} ({\rm cm}^{-3}\cdot {\rm s}^{-1}) \ ,\label{frrfktb}
\end{eqnarray}
\end{mathletters}
where 
$ \xi^{\rm (KT)} \equiv (\mu_K+\mu_p-\mu_n)/T$ 
and $I^{\rm (KT)}(u,T)$ is the integral over the kaon momentum divided by $T$, 
with $x=|{\bf p}_K|/T$: 
\begin{equation}
 I^{\rm (KT)}(u,T)
\equiv {1\over 6}\int_0^\infty dx
\frac{x^4}{(\widetilde{\omega_-}(x)+\mu_K/T)^3}
\frac{(\widetilde{\omega_-}(x)+u)}{1-e^{-\widetilde{\omega_-}(x)}}
\frac{[(\widetilde{\omega_-}(x)+u)^2+4\pi^2]}{
e^{\widetilde{\omega_-}(x)+u}-1} 
\label{ikt}
\end{equation}
where $\widetilde{\omega_-}(x)\equiv(\omega_-(x)-{\mu_K})/T $. 
In (\ref{frrfkta}), 
$g_A$(=1.25) is the axial-vector coupling strength, 
$\widetilde{f}\equiv f_{\pi NN}/m_\pi$ the $\pi NN$ 
coupling strength divided by the pion mass, 
$G_F$ the Fermi coupling constant, and $\theta_C$($\simeq$0.24) the Cabibbo 
angle. The factor $\displaystyle g_AG_F f\sin\theta_C\cos\theta_C
\cos^3\frac{\theta}{2}$ originates from a
$npK^-$ vertex factor in the transition matrix elements 
obtained from the chirally-rotated hadron currents,  
whereas the other factor, 
$\displaystyle{\widetilde f}^2 \frac{|{\bf p}_F(n)|^2}{|{\bf
p}_F(n)|^2+m_\pi^2}$ with ${\bf p}_F(i)$ ($i=n,p$) the Fermi momentum,
comes from the one-pion exchange potential introduced to describe the
long-range part of the interactions.  Due to the approximate
kinematical condition\cite{mti99-1}, only the nucleon axial-vector current
 (proportional to $g_A$) is found to contribute to the 
KT reactions after all the matrix elements for the 
lowest-order diagrams are summed. Hence the current for the $s$-wave
condensate,  which has only a time component, 
does not couple to the nonrelativistic  nucleon current. 
The remaining factor in (\ref{frrfkta}) arises from the phase space integrals 
in which $m_N^\ast$ is the effective nucleon mass and 
$T_9$, the temperature in units of 10$^9$ K. 
For simplicity, we take the values of the nucleon effective
 masses to be $m_p^\ast/m_N=m_n^\ast/m_N$=0.8\cite{fm79}.
 
For the forward KU process (KU-F) [(\ref{fku})],  one obtains
\begin{eqnarray}
\Gamma^{({\rm KU-F})}(\xi^{({\rm KU})},T) 
&=&\frac{G_F^2}{64\pi^5}\sin^2\theta_C\sin^2\theta\Big\{
10+3(g_A^2+9{\widetilde g_A}^2)\Big\}m_N^{\ast 
2}\mu_eT^5I_2(\xi^{({\rm KU})}) \cr
&=&(6.6\times 10^{29})\bigg({m^\ast_N \over m_N}\bigg)^2
{\mu_e\over m_\pi}\sin^2\theta(t)T_9^5I_2(\xi^{({\rm KU})}) \
({\rm cm}^{-3}\cdot {\rm s}^{-1}) 
\label{rrfku}
\end{eqnarray}
where $\widetilde g_A=F-\frac{1}{3}D$ = 0.15 with $F+D=g_A=1.25$,  
$D/(D+F)=0.658$\cite{fmtt94}, 
$\displaystyle I_2(u)\equiv \int_0 ^\infty dx x^2[\pi^2+(x+u)^2]/
( 1+\exp(x+u))$ and   
$\xi^{\rm (KU)} \equiv (\mu_e -\mu_K )/T$. 
That the rate for KU-F is proportional
to $\sin^2\theta$ is derived from the fact that the matrix element for 
KU-F is proportional to
$\sin\theta$. As a consequence, the KU reactions are operative only when a
condensate is present. 

For the forward MU process (MU-F) [(\ref{fmu})], we refer to Haensel's
result\cite{h92}: 
\begin{equation}
\Gamma^{({\rm MU-F})}(\xi^{({\rm MU})},T)
=(5.9\times 10^{23})\bigg({n_e\over n_0}\bigg)^{1/3}
\cos^2{\theta\over 2}T_9^7 J_2(\xi^{({\rm MU})})\ 
({\rm cm}^{-3}\cdot {\rm s}^{-1}) \ ,
\label{rrfmu}
\end{equation}
where 
$\displaystyle J_2(u)\equiv \int_0 ^\infty dx
x^2[9\pi^4+10\pi^2(x+u)^2+(x+u)^4]
/( 1+\exp(x+u))$, and 
$ \xi^{\rm (MU)}\equiv (\mu_p+\mu_e-\mu_n)/T$. 
Here, the matrix elements have been slightly modified 
by inclusion of an additional factor of $\cos^2(\theta/2)$ 
due to the presence of a condensate (cf. Paper I).  

The backward processes, denoted by the suffix B, are related to the forward 
processes by way of the following relations: 
\begin{equation} 
\Gamma^{\rm (KT-B)}(\xi^{\rm (KT)},T)
=e^{\xi^{\rm (KT)}}\Gamma^{\rm (KT-F)}(\xi^{\rm(KT)},T) \ , 
\label{brrkt}
\end{equation}
and 
\begin{equation} 
\Gamma^{\rm ({\rm KU-B})}(\xi^{({\rm KU})},T)
=\Gamma^{({\rm KU-F})}(-\xi^{({\rm KU})},T) \ , \ 
\Gamma^{\rm ({\rm MU-B})}(\xi^{({\rm MU})},T)  =\Gamma^{\rm ({\rm
MU-F})}(-\xi^{({\rm MU})},T) 
\label{brrkumu}
\end{equation} 
which are valid at low temperatures\cite{mti99-1}. 
The relation (\ref{brrkt}) between the forward and backward reaction rates 
for KT (where bosons are involved) 
differs from the corresponding relations (\ref{brrkumu}) 
in the case of the KU or MU process (where only fermions are involved). 

\subsection{Initial conditions}
\label{subsec:condition}

We adopt the following initial conditions: 
At $t$=0, we assume normal neutron-star matter ($\theta=0$), 
composed of nonrelativistic protons ($p$), neutrons ($n$),  
and ultra-relativistic free electrons ($e^-$), which is charge neutral,
$n_p^0=n_e^0$.
\footnote{The superscript `0'  denotes the  initial value at $t=0$.}
The baryon number density $n_B$ for the noncondensed state is taken to be  
larger than the critical density for kaon condensation, $n_B^C$ (=3$-$4 $n_0$). 
Initially, the system is assumed to be in $\beta$-equilibrium, i.e.,
$\mu_n^0=\mu_p^0+\mu_e^0$ [ $\xi^{({\rm MU})}(t=0)=0$] due to the rapid
$\beta$-decay reactions, $ n\rightarrow p+e^-+\bar 
\nu_e$, $ p+e^-\rightarrow n+\nu_e $, when the neutron star is hot at an early
stage.  The initial values for the proton fraction  
$x_p^0\equiv n_p^0/n_B$
and for the electron chemical potential $\mu_e^0$ are then obtained from the charge
neutrality condition, $n_Bx_p^0=(\mu_e^0)^3/(3\pi^2)$, and the
$\beta$-equilibrium condition where Eq.(\ref{ceppn}) is used with $\theta=0$. 
Finally, the total strangeness is assumed to be almost zero corresponding to
equal numbers of thermal $K^+$'s and
$K^-$'s present,  which, using Eqs. (\ref{cenergy}), (\ref{nkt}) and 
(\ref{eqdistribution}) with $\theta=0$, 
gives the initial value of the kaon chemical potential $\mu_K^0=-b^0$.  
As a consequence of these conditions, the initial kaon chemical potential 
$\mu_K^0$ has a large negative value, while $\mu_e^0$ is positive. 
Thus the system is far from chemical equilibrium 
with $|\xi^{({\rm KT})}(t=0)|=
|\xi^{({\rm KU})}(t=0)|=|(\mu_e^0-\mu_K^0)/T|\gg 1$. For example, 
one obtains $\xi^{({\rm KT})}(t=0)=-46$ and $\xi^{({\rm KU})}(t=0)$=46 
for $\Sigma_{KN}$= 300 MeV, $n_B$= 0.55 fm$^{-3}$ and $T=1\times
10^{11}$K. 
Starting from these initial conditions,
one can first obtain the number densities 
$n_i(t)$ at later times from the rate equations (\ref{rateeq}).  
Next one can obtain the electron chemical potential $\mu_e(t)$ from the relation
$\mu_e(t)=(3\pi^2n_e(t))^{1/3}$, the kaon chemical potential $\mu_K(t)$ and the
chiral angle $\theta(t)$ from Eqs.(\ref{nk})$-$(\ref{eqdistribution}) 
and (\ref{fieldeq}), and then the difference between $\mu_p(t)$ and $\mu_n(t)$ 
from Eq.(\ref{ceppn}). 
Because the charge neutrality condition, 
$ n_p(t)=n_K(t)+n_e(t) $, is built into 
 the rate equations (\ref{rateeq}), it can be seen that this condition is 
always satisfied.  The system
evolves dynamically  toward an equilibrated kaon-condensed phase 
through the nonequilibrium weak processes, KU, MU, and KT. 

\section{Numerical results and discussion}
\label{sec:result}

We choose the value for the $KN$ sigma term to be
$\Sigma_{KN}$=300 MeV\cite{dn86}. 
The critical density $n_B^C$ is then estimated to be $n_B^C$=0.49 fm$^{-3}$
(= 3.0 $n_0$ ) [c.f. Ref. \cite{yt99}]. In Table \ref{tabcond}, we list the values for
$\mu_K^0$ and  the proton fraction $x_p^0$ 
for the initial noncondensed state ($t$=0), 
and those of $\theta$, $\mu_K$ and $x_p$ for the kaon-condensed state in
chemical equilibrium ($t\rightarrow\infty$) [which we denote by 
the superscript `eq']. 
It has been shown in Ref.\cite{yt99} that the temperature dependence of 
the critical density and 
physical quantities in the kaon-condensed phase is weak 
for the temperature less than several tens of MeV.
Thus all the values listed in Table \ref{tabcond} have been estimated at $T$=0.  
Here, we study
the temporal behavior of the physical quantities  for two different baryon number
densities, 
$n_B$=0.55 fm$^{-3}$ and 0.70 fm$^{-3}$. The density $n_B$=0.55 fm$^{-3}$ is
close to $n_B^C$, such that we have a rather weak condensed state 
with small $\theta^{\rm eq}$. For $n_B$=0.70 fm$^{-3}$,  on the other
hand,  we have a well-developed condensed state with a large order
parameter $\theta^{\rm eq}$.  
When the baryon number density is increased above the critical density for 
kaon condensation, the number density of negatively-charged kaons increases as the
condensate develops. Charge neutrality of the system is maintained 
by the increase in the proton number density. Consequently, the proton fraction
$x_p^{\rm eq}$ increases as the baryon number density increases 
in an equilibrated kaon-condensed state.  On the other hand,
the negatively charged electrons are replaced  by the negatively charged kaons 
as a condensate develops with an increase in the baryon number density.   
The consequent decrease in the electron abundance produces a decrease 
in the charge chemical potential
$\mu_e^{\rm eq}$ (=$\mu_K^{\rm eq}$), resulting in a large proton fraction and
a reduced charge chemical potential, which are two characteristic features of
 the kaon-condensed state\cite{t95,fmmt96,l96,pbpelk97}.  

\subsection{Two typical time scales}
\label{subsec:times}

\begin{figure}[tt]
\noindent\begin{minipage}[l]{0.5\textwidth}
\centerline{
\epsfxsize=\textwidth
\epsffile{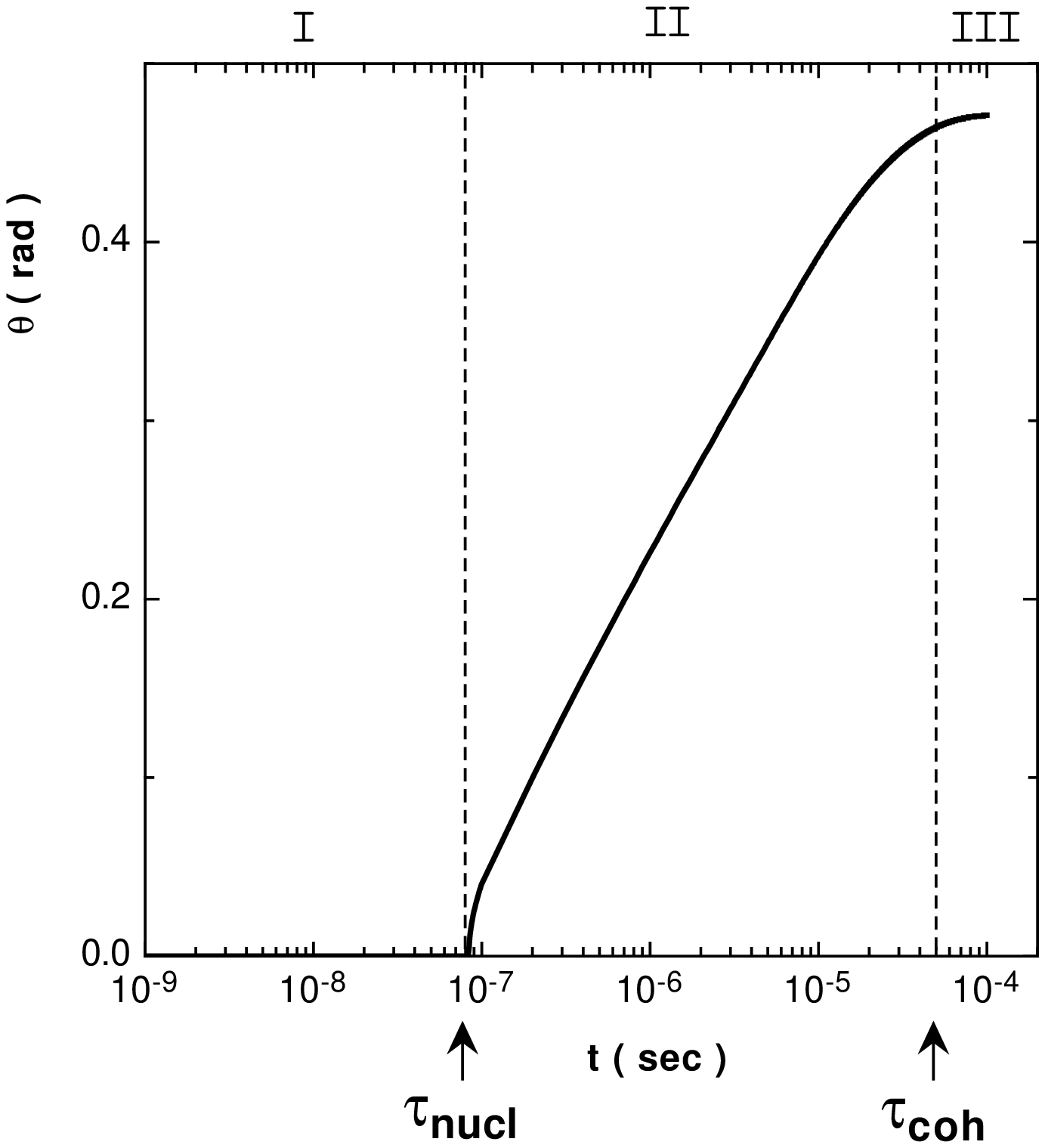}}
\end{minipage}~
\begin{minipage}[r]{0.5\textwidth}
\centerline{
\epsfxsize=\textwidth
\epsffile{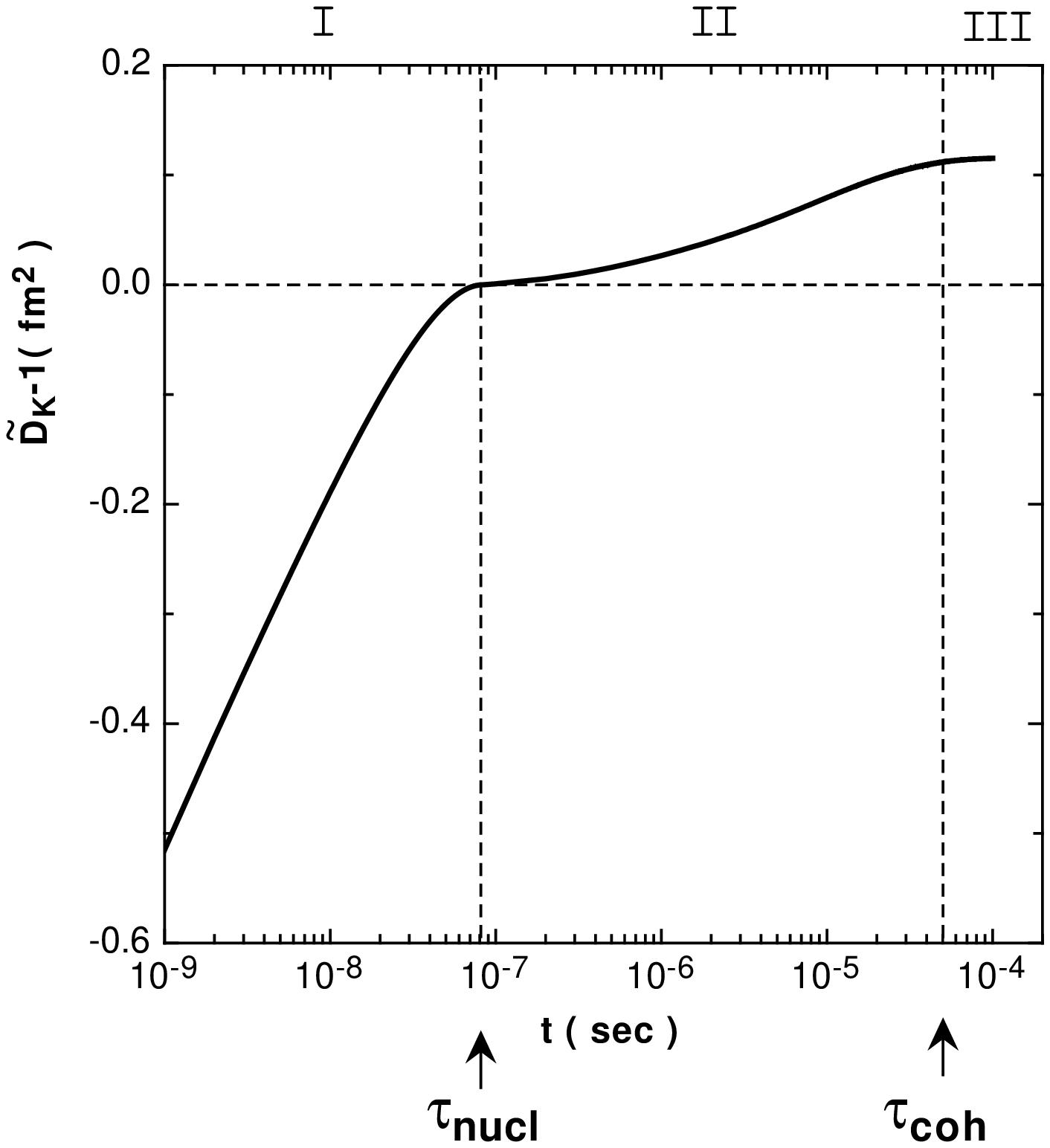}}
\end{minipage}
\caption{ (a) Chiral angle $\theta$ and (b) $\widetilde D_K^{-1}$, 
the negative of 
the expansion coefficient for the thermodynamic potential per unit volume 
$\Omega/V$ with respect to the squared classical kaon field, as a
function of time  for the baryon number density
$n_{\rm B}$=0.55 fm$^{-3}$ and the temperature
$T$ =1.0$\times 10^{11}$ K. }
\label{fig1} 
\end{figure}

In Fig.1 (a), we show the chiral angle $\theta$ 
as a function of time for $n_{B}$=0.55 fm$^{-3}$  
and $T$=1.0$\times 10^{11}$ K. For the initial values, 
$\mu_K^0=-b^0=-139$ MeV and
$x_p^0$=0.14. The evolution of the
kaon condensate may be divided into the following three stages: 
(I) no condensate ($\theta(t)=0$) with thermal kaons present   
until the onset of condensation, 
(II)  onset of condensation and its monotonic growth,  and (III) the asymptotic
stage near chemical equilibrium. In equilibrium, one finds $\theta^{\rm
eq}$=0.48 (rad) from Fig. \ref{fig1}(a), and the kaon chemical potential 
and the proton fraction have the values $\mu_K^{\rm eq}$=203 MeV and 
$x_p^{\rm eq}\equiv n_p^{\rm eq}/n_{\rm B}=0.23$, respectively
 (see Table \ref{tabcond}).

\begin{figure}[t]
\centerline{
\epsfxsize=0.5\textwidth\epsffile{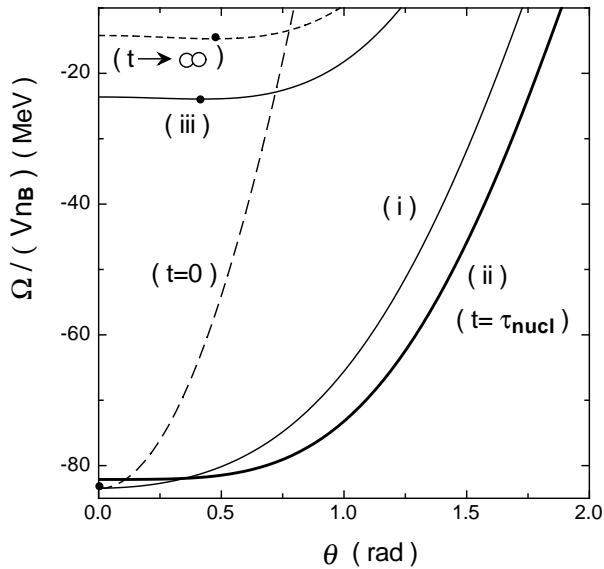}}
\caption{\footnotesize Thermodynamic potential per particle  
$\Omega/(Vn_B)$ as a function of $\theta$ corresponding 
to the three cases (i)$-$(iii). 
The long-dashed and short-dashed lines correspond to the cases 
at $t=0$ and $t\rightarrow\infty$, respectively. 
The dots on the lines represent the locations 
of the minima of $\Omega$. }
\label{fig2}
\end{figure}

The onset of condensation from a noncondensed state and its subsequent 
growth are related to the change in shape of 
the thermodynamic potential $\Omega$ as a function of $\theta$ around
$\theta=0$.  This change in shape is also associated with the appearance  
of a soft kaon mode, reflected in 
a singularity of the distribution function $f_K({\bf p}_K)$,
Eq.(\ref{eqdistribution}). 
To see  this, we expand $\Omega(t)/V$
around
$\theta=0$: 
\begin{equation}
\Omega(t)/V=-{f^2\over 2}\widetilde D_K^{-1}\theta(t)^2
+O(\theta(t)^4) \ ,
\label{eqexpansion}
\end{equation}
with $\widetilde D_K^{-1}\equiv D_K^{-1}\big(\omega=\mu_K(t),{\bf
p}_K=0;n_B\big)$, where $D_K^{-1}\big(\omega,{\bf p}_K;n_B\big)$ is the 
inverse kaon propagator:
\begin{equation}
D_K^{-1}\big(\omega,{\bf p}_K;n_B\big)\equiv \omega^2-{\bf p}_K^2
+2b(t)\omega-m_K^{\ast 2} \ .
\label{eqdkinv}
\end{equation}
Beginning with
$\mu_K=-b^0$  at $t$=0, the kaon chemical potential $\mu_K(t)$ 
increases monotonically with time. As a result,  
the factor $\widetilde D_K^{-1}$ changes sign, depending on the value of
$\mu_K$ relative to the critical value, 
$\mu_K^C(t)\equiv -b(t)+\Big(b(t)^2+ m_K^{\ast 2}\Big)^{1/2}$,  
defined as a root of $\widetilde D_K^{-1}$.  
In Fig. \ref{fig2}, we show the
thermodynamic potential per particle $\Omega/(Vn_B)$ 
as a function of $\theta$ for $n_{\rm B}$=0.55 fm$^{-3}$ and $T$ =1.0$\times
10^{11}$ K and the other parameters $n_i$ and 
$\mu_i$ ($i=p,n,e^-,K^-$) fixed at three times corresponding to cases (i)$-$(iii).  
 Case (i) corresponds to a time before the onset of condensation, for which 
$\mu_K$ is smaller than $\mu_K^C$. [E.g., at $t$=0,
$\mu_K=-b^0<\mu_K^C(t=0)$.] In this case, 
$\widetilde D_K^{-1}$ is negative, 
which means that 
the thermodynamic potential per unit volume 
$\Omega/V$ is convex at $\theta$=0, as seen from Eq. (\ref{eqexpansion}) 
and the curve (i) in Fig. \ref{fig2}, 
and the noncondensed state ($\theta$=0) is stable against fluctuations in 
$\theta$. 
Because $\mu_K^C$ corresponds to the lowest excitation energy 
of the $K^-$,  
$\omega_-({\bf p}_K=0)$ {\it in normal matter},  
 one can see that
$\omega_- ({\bf p}_K) > \omega_-({\bf p}_K=0)=\mu_K^C>\mu_K$. 
Thus the Bose-Einstein distribution function 
$f_K$ here has no singularity. Case (ii) corresponds to a time for which 
$\mu_K(t)=\mu_K^C(t)$, where 
$\widetilde D_K^{-1}=0$.   In this case, the system becomes unstable with respect to
the fluctuations in $\theta$ (the curve (ii) in Fig. \ref{fig2}), 
and since $\omega_-({\bf p}_K=0)$($=\mu_K^C$) is equal to $\mu_K$,
$f_K({\bf p}_K)$ becomes divergent at ${\bf p}_K$=0. Thus the appearance of a
soft kaon mode with  
$\omega_-({\bf p}_K=0)$=$\mu_K$ implies the  
onset of an instability due to the formation of a condensate. 
Finally, case (iii) corresponds to a time after the onset of condensation, for which 
$\mu_K(t)$exceeds 
$\mu_K^C(t)$, so that $\widetilde D_K^{-1}>0$. In this case, 
$\Omega/V$ is concave near $\theta$=0 as a function of $\theta$;  
which means that the noncondensed  state
corresponds to the maximum of  $\Omega/V$, as seen from the curve (iii) in
Fig. \ref{fig2}, and is unstable   against the formation of a condensate. 
The system therefore evolves into a  condensate described by 
$\theta(t)$, which is  determined adiabatically by a minimum 
in the thermodynamic potential [cf. Eq.(\ref{fieldeq})].  
By the use of the dispersion relations for the kaon  {\it in the
condensed phase} (see \ref{subsec:thermal})\cite{ty98},  it can 
be shown that the lowest excitation energy of $K^-$ 
{\it in the condensed phase} is equal to 
$\mu_K$, i.e., $\omega_-({\bf p}_K=0)=\mu_K(t)$; there is a soft kaon
mode in the condensed phase, and $f_{K}({\bf p}_K)$ is divergent 
at ${\bf p}_K=0$, as in the case (ii).  From Fig. 1(b), where 
$\widetilde D_K^{-1}$ is shown  as a function of time, one sees that the  function
$\widetilde D_K^{-1}$  changes sign at the onset of condensation. 
We can call the onset time for condensation {\it the nucleation time}, 
$\tau_{\rm nucl}$, and the time for growth of 
the condensate {\it the coherence time}, $\tau_{\rm coh}$, after  Stoof's
discussion about the formation of BEC in an atomic gas
\cite{gss95}.  From Fig. \ref{fig1}, one finds $\tau_{\rm nucl}=8\times 10^{-8}$
sec,  and $\tau_{\rm coh}=5\times 10^{-5}$ sec for $n_{\rm B}$=0.55
fm$^{-3}$ and $T=1.0\times 10^{11}$ K.   

\subsection{Connection between the reaction rates and the change in chemical
composition}
\label{sub:react}

\begin{figure}[t]
\centerline{
\epsfxsize=0.5\textwidth\epsffile{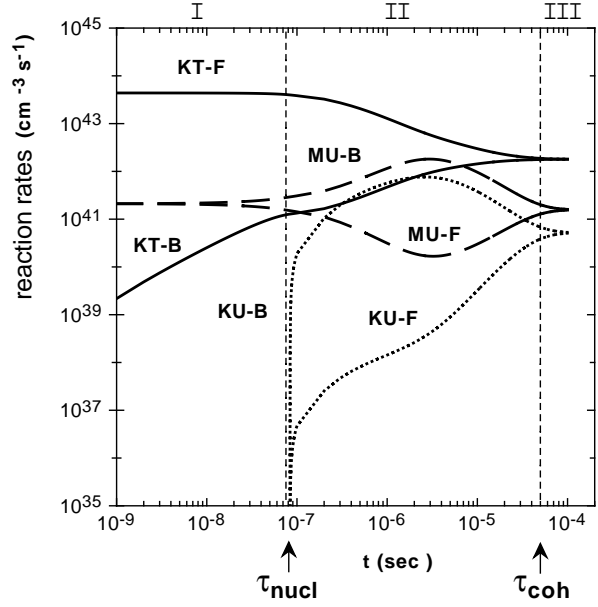}}
\caption{\footnotesize Temporal behavior of the reaction rates for the weak
reactions, KT, KU, and MU for $n_{\rm B}$=0.55 fm$^{-3}$ and
$T=1.0\times 10^{11}$ K. }
\label{fig3}
\end{figure}

 The temporal behavior of the reaction rate for each weak
process is shown for $n_{\rm B}$=0.55 fm$^{-3}$ and
$T=1.0\times 10^{11}$ K in Fig. \ref{fig3}.  
The solid,  dashed and 
dotted lines denote the KT, MU, and KU processes, respectively. 
The curves show that the forward thermal kaon process KT-F is  
dominant in magnitude over the KU and MU reactions 
throughout the equilibration process\cite{mti99-1}. 
Hence the production of the thermal and condensed kaons 
proceeds mainly via the KT-F reaction,  which is responsible for the onset of 
condensation and its subsequent growth. 

\begin{figure}[t]
\centerline{
\epsfxsize=0.5\textwidth\epsffile{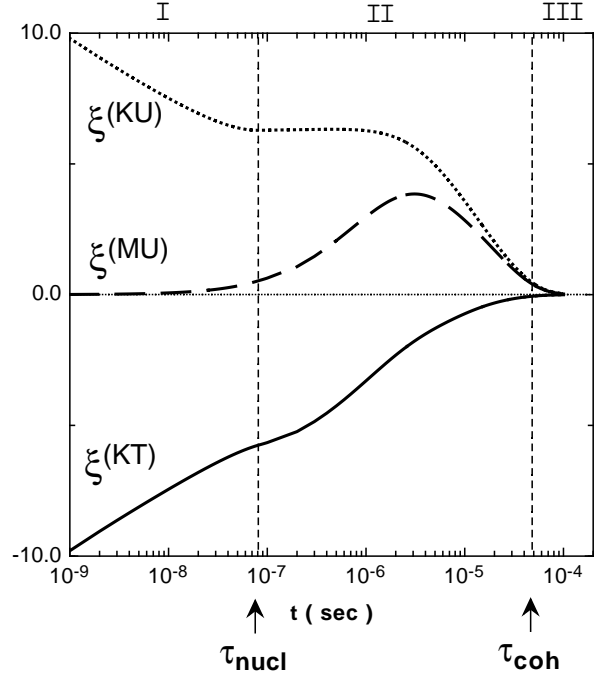}}
\caption{\footnotesize Temporal behavior of the dimensionless
parameters $\xi^{(\alpha)}$ ($\alpha$= KU, MU, KT) for $n_{\rm B}$=0.55
fm$^{-3}$ and $T=1.0\times 10^{11}$ K. }
\label{fig4}
\end{figure}

In Fig. \ref{fig4}, we show the temporal behavior of the dimensionless
parameters $\xi^{(\alpha)}$ ($\alpha$= KU, MU, KT),
which measure the deviation from chemical equilibrium. 
 Beginning with a large negative value, 
$\xi^{({\rm KT})}$ increases monotonically toward zero, 
where the system is in chemical equilibrium with respect to 
the KT reactions. Note that the KT reaction rates are large even in the initial
stage where the system is far from chemical equilibrium ($|\xi^{({\rm KT})}(t)|
\gg1$). In Paper I, we examined in detail the characteristic roles of the
soft and hard thermal kaons in the KT reactions,  and demonstrated that 
the high-energy component of the thermal kaons  contributes to 
the KT reaction rates during the noncondensed stage.
On the other hand, a low energy component (a soft mode ) of the thermal kaon
excitations in the kaon-condensed state was shown to contribute mainly 
near chemical equilibrium\cite{mti99-1}. 

\begin{figure}[t]
\centerline{
\epsfxsize=0.5\textwidth\epsffile{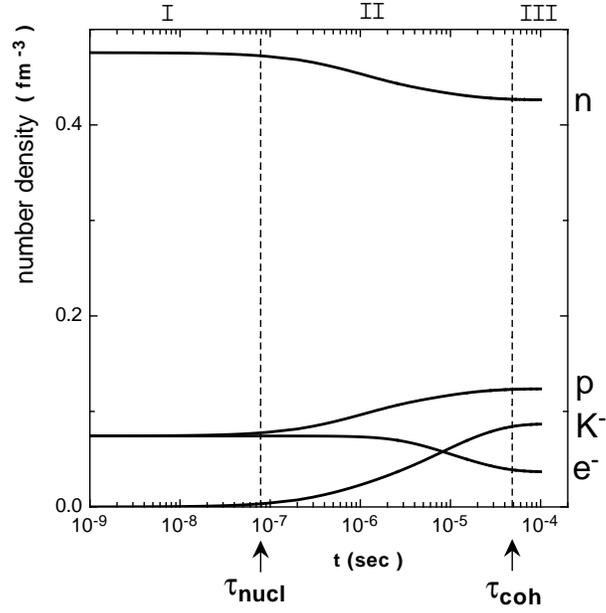}}
\caption{\footnotesize Temporal behavior of the number densities 
of the chemical species. }
\label{fig5}
\end{figure}

Let us now examine how these reactions change the chemical composition. 
The number densities of the chemical species are shown in Fig. \ref{fig5} for the
same density and temperature 
as in Figs. \ref{fig1}$-$\ref{fig4}. In stage I ($0<t<\tau_{\rm
nucl}$), where there is no condensate, the kaons produced 
by the KT reactions are thermal, and the KU reactions cannot proceed  because there
is no condensate ($\theta=0$). The initial
$\beta$-equilibrium is disturbed by the proton excess, created
through  the KT-F reaction, and the MU reactions are enhanced because  
the MU-B reaction proceeds more rapidly than the MU-F reaction, which 
restores $\beta$-equilibrium by reducing the proton excess. 
Nevertheless, the
proton number density $n_p$ still increases, 
since the KT-F reaction rate is larger 
than the MU-B reaction rate, as can be seen in Fig. \ref{fig3} (see also Paper I). 
Because the total baryon number density 
$n_{\rm B}$ is constant, the neutron number density $n_n$ therefore 
decreases with
time.  In stage I, the change in the electron number density $n_e$ is brought about
only by the MU reactions, and $n_e$ decreases slightly with time, 
the MU-B reaction proceeding more rapidly than the MU-F reaction. 
The extent of the changes in $n_p$, $n_n$, and $n_e$ is  very small. 
Consequently, although 
the electron chemical potential $\mu_e$ decreases with time because of 
the decrease in $n_e$, this potential, along with the proton and the neutron 
chemical potentials $\mu_p$ and $\mu_n$ remain almost constant in stage I.  
On the other hand, the kaon chemical potential
$\mu_K$ increases  significantly with time; $\mu_K^0=-b^0$ at $t$=0 and 
$\mu_K\simeq\mu_K^{\rm eq}$ at
$t=\tau_{\rm nucl}$. Hence $\xi^{({\rm KU})}$, which is
proportional to
$\mu_e-\mu_K$, decreases monotonically with time and
$\xi^{({\rm KT})}$ increases monotonically with time [cf. Fig. \ref{fig4}],  while
$\xi^{({\rm MU})}$ remains close to its initial value ($\sim 0$) 
 except for the time near $\tau_{\rm nucl}$. 

In stage II ($\tau_{\rm nucl}<t<\tau_{\rm coh}$), the KU reactions can proceed 
due to the appearance of a condensate, but the qualitative behavior of each
chemical composition is similar to stage I. In particular, 
the kaon and  proton number densities   
increase through the KT reactions, while those of the electrons and neutrons 
decrease. The change in the number density of the kaons $n_K$ 
comes mainly from  the condensed part $\zeta_K$ [(\ref{nk})] 
after the saturation of the thermal part $n_K^T$, 
with the condensed part $\zeta_K$ roughly
proportional to the square of $\theta$, $\zeta_K\propto\theta^2$, 
for a small chiral angle, 
as seen from Eq. (\ref{nkc}). According to Fig.\ref{fig1}(a), the chiral angle
$\theta$ increases rapidly with time for $t\gtrsim\tau_{\rm nucl}$. 
Thus the change in the number density of the kaons  
$n_K$ is substantial, being more pronounced in stage II than in stage I 
due to the appearance of a condensate. 
The change in the number density of the electrons $n_e$ in stage II is caused 
by the KU and MU reactions. But since the reaction rates for the electron absorption
processes, KU-B and MU-B, are smaller by orders of magnitude 
than the rate for the kaon production process KT-F, 
 the change in the number density of the electrons is less marked 
than the change in the number density of the kaons, and remains 
almost constant near the onset of condensation, until both the MU-B and KU-B
reactions become maximum and operate significantly 
to reduce $n_e$ in the later part of
stage II (cf. Figs. \ref{fig3} and \ref{fig5}). From the temporal behaviors of
$n_K$, $n_e$, the charge neutrality, 
$n_p=n_e+n_K$, and baryon number conservation, $n_p+n_n=n_B$, 
it can be seen that the change in the number densities of the protons 
and neutrons become remarkable in stage II. 

During the earlier part of stage II where the number density of the electrons is 
almost constant, the parameter
$\xi^{({\rm KU})}$ remains unchanged because both the electron
and the kaon chemical potentials are almost constant. On the other hand, 
as a result of the increase in the proton number density and the decrease
in the neutron number density, the difference between the proton and neutron
chemical potentials becomes larger with time, 
as seen from Eq.(\ref{ceppn}).
Thus the parameter $\xi^{({\rm MU})}$ increases with time, after which 
 both $\xi^{({\rm KU})}$ and $\xi^{({\rm MU})}$ decrease 
with time, and the system enters into the final stage III
(cf. Fig. \ref{fig4}). 

The coherence time
$\tau_{\rm coh}$ may be identified with the time 
at which the magnitudes of the forward and backward reaction rates for KT  
become equal. As seen in Fig.\ref{fig3}, chemical equilibrium 
with respect to the KU and MU reactions is achieved 
at a later time than that at which the KT reaction reaches equilibrium.  

Finally, in stage III ($t>\tau_{\rm coh}$), where the system is 
close to chemical equilibrium ($\xi^{({\rm KT})}\simeq 0$), 
 both the KU and MU reactions compete with each other to determine the
dynamical behavior of the system. 
In this stage, the deviation parameters $\xi^{({\rm KU})}$ and
$\xi^{({\rm MU})}$ damp exponentially with time, and the system
 approaches chemical equilibrium. 
The relaxation time $\tau_{\rm rel}$ can be calculated analytically 
[see Eq. (\ref{atau}) in Appendix A].  
For $n_{\rm B}$=0.55 fm$^{-3}$ and $T$=1.0$\times 10^{11}$ K, 
the analytic result gives $\tau_{\rm rel}=2\times 10^{-4}$ sec,  
which is in agreement with the numerical result read from Fig. \ref{fig3}.  

\subsection{Temperature-dependence of the characteristic time scales }
\label{sub:temp}

\begin{figure}[t]
\centerline{
\epsfxsize=0.5\textwidth\epsffile{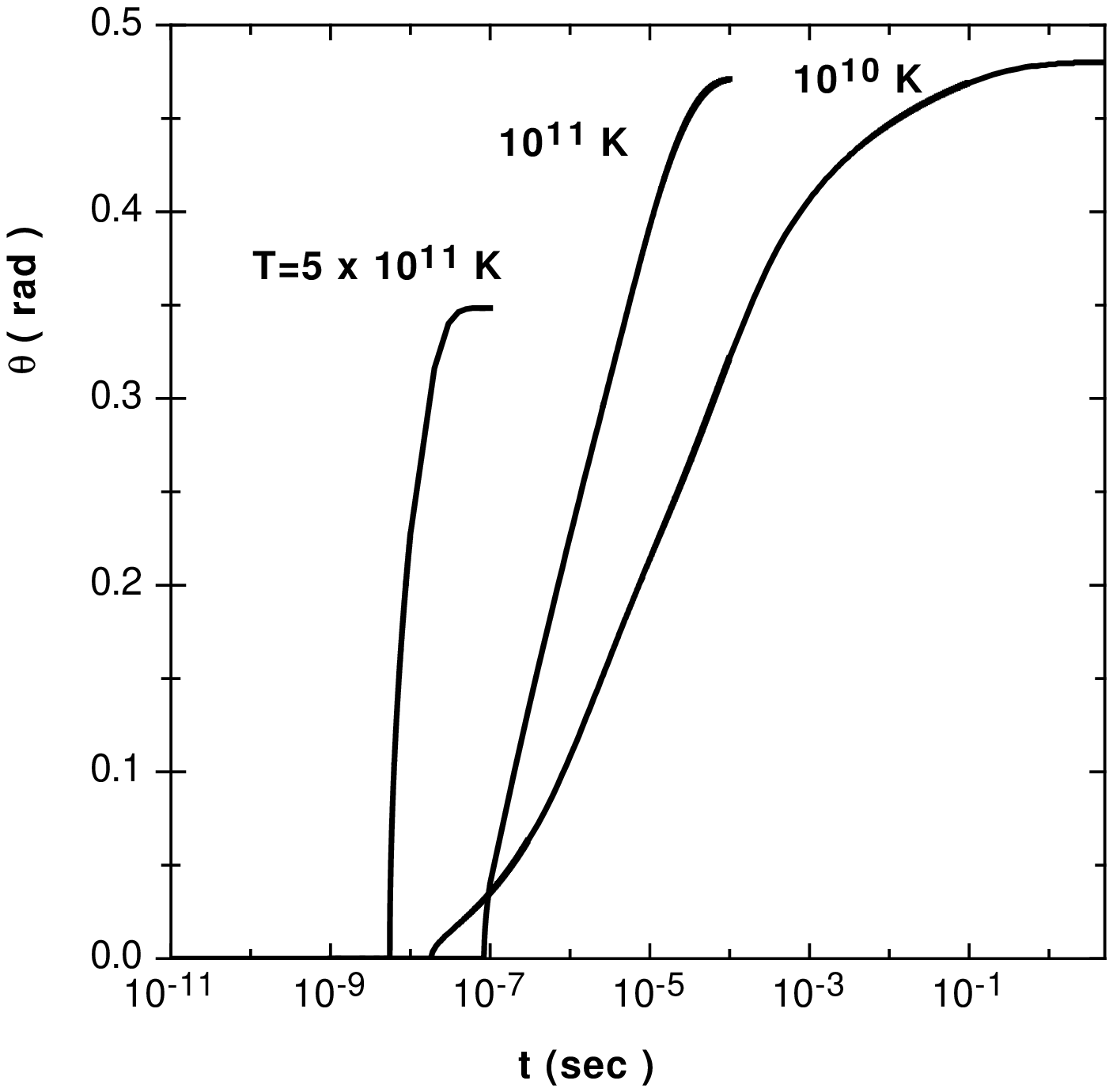}}
\caption{\footnotesize Temporal behavior of the  chiral
angle $\theta$ for $n_{\rm B}$=0.55 fm$^{-3}$ and three different temperatures,
$T=1.0\times 10^{10}$ K, $1.0\times 10^{11}$ K, and $5.0\times 10^{11}$ K.}
\label{fig6}
\end{figure}

Next we compare the temperature dependence of the characteristic time
scales, $\tau_{\rm nucl}$, $\tau_{\rm coh}$, and $\tau_{\rm rel}$ 
at a fixed baryon number density. 
Figure \ref{fig6} shows the behavior of the chiral
angle $\theta$ for $n_{\rm B}$=0.55 fm$^{-3}$ 
at three different temperatures,
$T=1.0\times 10^{10}$ K, $1.0\times 10^{11}$ K, and $5.0\times 10^{11}$ K. 
In addition, Table \ref{tabtau} lists the values of $\tau_{\rm nucl}$ and
$\tau_{\rm coh}$, estimated from the 
numerical calculations, and the value of $\tau_{\rm rel}$, obtained from
Eq. (\ref{atau}). 
 Roughly speaking, 
it can be seen that $\tau_{\rm nucl}$ depends weakly on temperature, 
while $\tau_{\rm coh}$ and $\tau_{\rm rel}$ depend sensitively on temperature: 
In particular, from Table \ref{tabtau} and Fig. \ref{fig6},  
one obtains $\tau_{\rm nucl}$=$2\times10^{-8}$ sec  
$\rightarrow$ $6\times 10^{-9}$ sec,  
 $\tau_{\rm coh}$=$2\times10^{-1}$ sec
$\rightarrow $ $4\times 10^{-8}$ sec, and $\tau_{\rm rel}$=40 sec
$\rightarrow$ $5\times 10^{-7}$ sec as $T=1.0\times 10^{10}$ K 
$\rightarrow$ $5.0\times 10^{11}$ K for $n_B$=0.55 fm$^{-3}$. 

Consider first the temperature 
dependence of the nucleation time, $\tau_{\rm nucl}$, which provides a typical
time scale for saturation of the thermal part of the strangeness number 
density $n_K^T(t)$. For $t>\tau_{\rm nucl}$,  $n_K^T(t)$ changes little. In chemical
equilibrium,  the thermal kaons occupy a progressively larger part of  the total
strangeness density as the temperature increases, and two competing effects
produce the temperature dependence of
$\tau_{\rm nucl}$.  (i) The value of $n_K^T(t)$ at saturation  
 is larger at higher temperature, 
which tends to increase the nucleation time, $\tau_{\rm nucl}$. 
On the other hand, (ii) a higher temperature gives a higher reaction rate 
for the relevant KT process. (cf. Paper I). As a consequence  
the thermal kaon number density $n_K^T(t)$ comes to saturation earlier.

In contrast, the temperature dependence of the coherence time 
$\tau_{\rm coh}$ is explained by the fact that after a condensate appears, its
subsequent growth in stage II is  controlled mainly by the KT reactions,  
as seen in Sec. \ref{sub:react}, and the KT reaction rates depend sensitively 
on the temperature over the entire nonequilibrium processes, 
(cf. Paper I). This leads to a sensitive temperature dependence 
of $\tau_{\rm coh}$. 

Finally, the relaxation time $\tau_{\rm rel}$ depends on the 
KU and MU reaction rates through the quantities 
$\displaystyle\widetilde\Gamma^{({\rm KU-F})} \ (\propto T^5)$ and 
$\displaystyle\widetilde\Gamma^{({\rm MU-F})} \ (\propto T^7)$ 
[see Eq. (\ref{tildegamma}) in Appendix A], and from Eqs. (\ref{atau}) and
(\ref{tildegamma}),  it can be seen that the relaxation time $\tau_{\rm rel}$
depends sensitively on the temperature, so as to
decrease significantly as the temperature increases. 

\subsection{Density-dependence of the characteristic time scales }
\label{sub:density}

\begin{figure}[t]
\centerline{
\epsfxsize=0.5\textwidth\epsffile{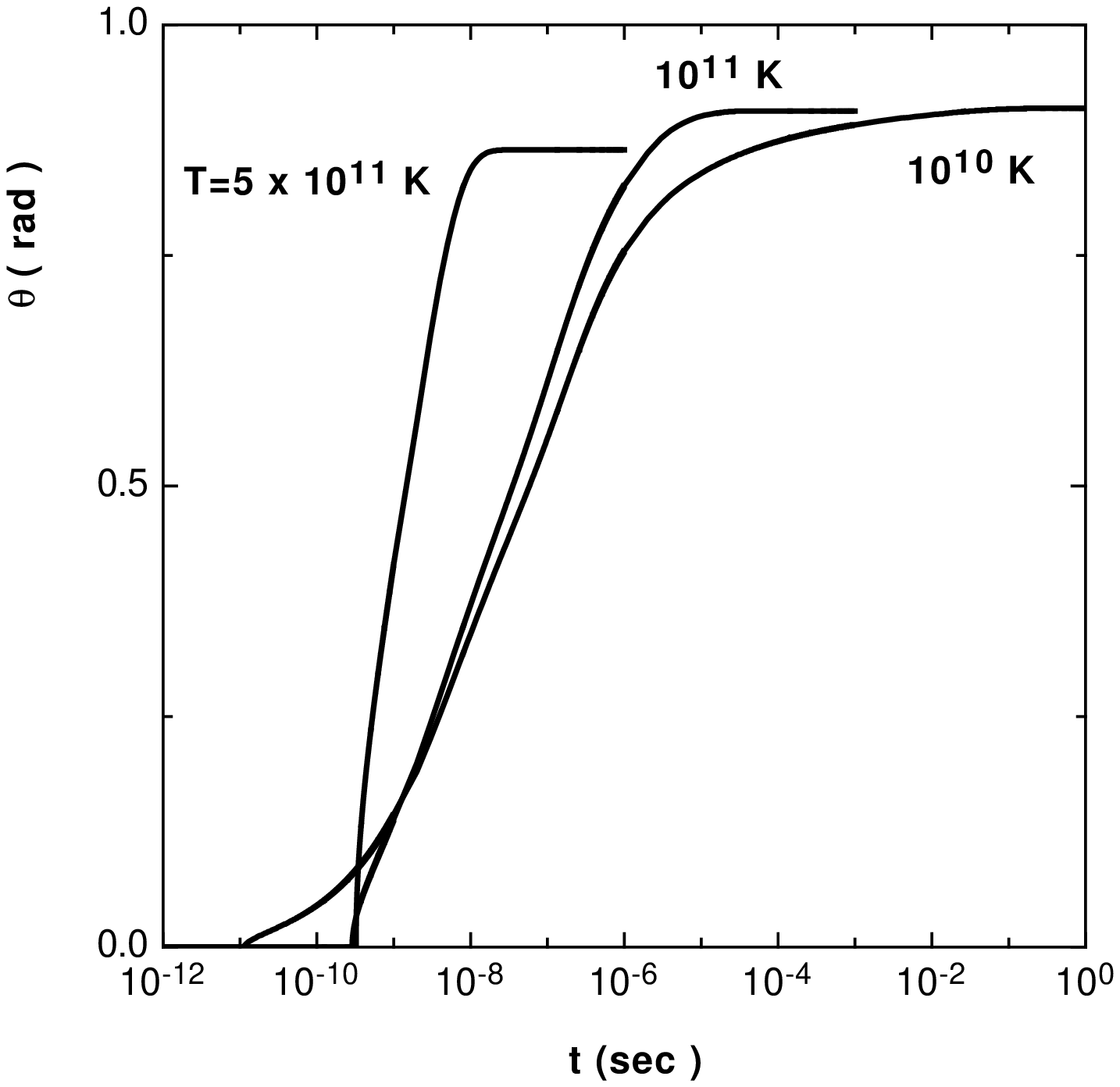}}
\caption{\footnotesize Temporal behavior of the  chiral
angle $\theta$ for $n_{\rm B}$=0.70 fm$^{-3}$ and three different temperatures,
$T=1.0\times 10^{10}$ K, $1.0\times 10^{11}$ K, and $5.0\times 10^{11}$ K.}
\label{fig7}
\end{figure}

 Here we discuss  the temporal behavior of the system 
at a higher baryon number density, $n_{\rm B}$=0.70 fm$^{-3}$, for which  
one can expect a fully developed kaon-condensed phase 
when the system is in chemical equilibrium.  
Initially we start with $n_K^T(t=0)=0$, which gives  
$\mu_K^0=-b^0=-180$ MeV and $x_p^0$=0.16. 
The temporal behavior of the  chiral angle $\theta$ is shown 
in Fig. \ref{fig7} for $n_{\rm B}$=0.70 fm$^{-3}$ for the same three temperatures
as in Fig. \ref{fig6}. 
The numerical values for the nucleation time $\tau_{\rm nucl}$, 
the coherence time $\tau_{\rm coh}$, and  
the relaxation time  $\tau_{\rm rel}$ for the density and temperatures 
corresponding to the three curves shown are listed in Table \ref{tabtau}. 
From an examination of this table and Fig. \ref{fig7}, it can be seen that while 
the nucleation time $\tau_{\rm nucl}$ depends weakly on the temperature,  
the coherence time $\tau_{\rm coh}$ and the relaxation time $\tau_{\rm rel}$ 
depend sensitively on the temperature, 
becoming shorter as the temperature is raised. 
These features for $\tau_{\rm nucl}$, 
$\tau_{coh}$, and $\tau_{\rm rel}$ are qualitatively the same
as in the case of the lower baryon number density $n_{\rm B}$= 0.55 fm$^{-3}$. 

The time scales obtained for $n_B$=0.70 fm$^{-3}$ can be compared 
with those for $n_B$=0.55 fm$^{-3}$.  
From Table \ref{tabtau}, one can see that the nucleation time 
$\tau_{\rm nucl}$ is sensitive to the baryon number density, 
with the ratio $\displaystyle \tau_{\rm nucl} \ (0.55 \ {\rm fm}^{-3})/
\tau_{\rm nucl} \ (0.70 \ {\rm fm}^{-3})$ equal to 
2$\times 10$ $-$ $2\times 10^3$ for $1\times 10^{10}$ K$<T< 5 \times
10^{11}$ K. This density dependence is explained as follows: 
In the noncondensed state where $|\xi^{({\rm KT})}|\gg 1$ ($\xi^{({\rm KT})}<0$), 
the KT-F reaction rate is very sensitive to $|\xi^{({\rm KT})}|$ and becomes large
rapidly as $|\xi^{({\rm KT})}|$ increases\cite{mti99-1}. 
In addition, the kaon chemical potential
$\mu_K$ ($<$0 at $t=0$) in $|\xi^{({\rm KT})}|$ increases rapidly with time in
stage I, and, before the onset of condensation, it is already almost equal to the
equilibrium value $\mu_K^{\rm eq}$ ($>$0), which has a smaller value for a larger
baryon number density. As compared with the density dependence of 
$\mu_K$, the density dependence of the difference 
between the proton and neutron chemical potentials 
$\mu_p-\mu_n$ ($<0$) in $|\xi^{({\rm KT})}|$ is weak. 
Hence, as the baryon number density increases, the parameter
$|\xi^{({\rm KT})}|$  (=$|\mu_K+\mu_p-\mu_n|/T\gg 1$) increases mainly due to
the decrease in $\mu_K$. As a result, the KT-F reaction rate 
in  stage I is much larger 
for a larger baryon number density, which considerablly reduces 
the time required for saturation of the thermal kaon number density. 

After the appearance of a condensate, the value of $|\xi^{({\rm KT})}|$
becomes sufficiently small, so that the density dependence of $|\xi^{({\rm KT})}|$ 
becomes insignificant. Therefore, the KT-F reaction rate does not depend much on
$n_B$. Hence the coherence time $\tau_{\rm coh}$, which is controlled  
by the KT-F reaction, shows only a weak dependence on $n_B$. In a similar
way, near chemical equilibrium where
$|\xi^{({\rm KU})}|\ll 1$ and 
$|\xi^{({\rm MU})}|\ll 1$, the density dependence of both the KU-F and MU-F 
reaction rates is weak, and therefore $\tau_{\rm rel}$, which is determined 
by these reaction rates, is found to have a weak dependence on $n_B$. 

\subsection{Implications for the delayed collapse of a neutron star}
\label{subsec:implications}

From Table
\ref{tabtau}, one can see that a condensate develops fully 
with a characteristic time
scale given by the coherence time
$\tau_{\rm coh}$. In the context of the delayed collapse 
of a newly born neutron star, we consider some implications of the results. 
Since the neutrino degeneracy is not taken into account in our framework, 
 our results are applicable to the cooling era 
 after the deleptonization. Hence the delayed collapse associated with
 a kaon condensate should be considered in this cooling era. 

The low temperature case (e.g., $T=1\times10^{10}$ K) may apply 
to the final stage of the cooling era, while   
 the high temperature case (e.g.,$T\gtrsim 10^{11}$ K) to 
the initial stage. In both cases, we have found that 
the time scale of $\tau_{\rm coh}$ is very small; 
$\tau_{\rm coh}\sim$ 0.1 sec for $T=1\times10^{10}$ K and 
 $\tau_{\rm coh}\lesssim 5\times 10^{-5}$ sec 
 for $T\gtrsim 10^{11}$ K, which should be compared with the cooling time scale 
of order of ten seconds. 
 Hence we can conclude that, in both cases,  the
time delay of a collapse due to the formation of a condensate is 
negligible as compared with the cooling time scale 
and will have a minor effect on the evolution of the
protoneutron star. 

\section{Summary and concluding remarks}
\label{sec:summary}

We have considered the  kinetics of kaon condensation by the use of rate
equations which include the three weak reactions: the thermal kaon process, 
the kaon-induced Urca process, and the modified Urca process. 
The thermal kaon process is shown to be dominant over 
other weak reactions throughout the equilibration process. 
The evolution of the kaon condensate is divided into the 
following three stages: (I) the noncondensed stage before the onset 
of condensation, 
(II) onset of condensation and its growth until its saturation,  
and (III) the asymptotic relaxation stage near chemical equilibrium. 
The role of thermal kaons and especially 
the connection between the existence of a soft kaon mode 
and the instability of the noncondensed state has been clarified. 
It has been found that a full development of a condensate is 
characterized by a time scale given by the coherence time $\tau_{\rm coh}$.
Using these results, we have made a brief comment on implications of 
our results on the delayed collapse mechanism of a newly-born neutron star 
into a black hole. The situation adopted in our framework 
may be applicable to the  cooling era after the deleptonization.
It is found that the time scale of the development of a condensate 
(the coherence time) is much smaller than  that of the cooling 
for the relevant temperatures (1 MeV$-$ several tens of MeV).
Therefore,  the time delay for the formation of a kaon condensate may not affect 
the delayed collapse of a neutron star. 

 Several effects which should be taken into account  within our framework in the future
are as follows:

\noindent (1) 
We have used the thermodynamic potential neglecting any fluctuations except 
for those produced by thermal kaons. However, 
for a neutron star just born in a supernova, since 
the initial temperature is as high as several tens of MeV,  
thermal fluctuations other than the kaon thermal loop 
contributions\cite{ty98} should be fully incorporated into the thermodynamic
potential. 

\noindent (2) 
Furthermore, since the matter is opaque to the neutrinos at high temperatures, 
the thermal effects associated with neutrino diffusion may also be important. 
In particular, the degenerate neutrinos will contribute to the thermodynamic
potential and  influence the relevant reaction rates. Therefore, 
the neutrino diffusion may affect the thermal evolution of 
 a newly-born neutron star.

\noindent (3) 
The reaction rates for the relevant weak processes have 
been obtained in the low temperature approximation (c.f. Paper I). 
The expressions for the reaction rates 
must be extended to the high temperature case, and the
 phase-space integrals
should be performed numerically beyond the low temperature 
approximation\cite{gps96}. 

\noindent (4) During the equilibration process, we have assumed that 
the temperature $T$ is constant: the energy 
produced by the weak reactions is lost to the surroundings. However,
because the energy release due to the nonequilibrium weak reactions 
will heat up the matter, the temperature is in general
time-dependent during the course of the equilibration process, 
and this can affect the evolution of the star. 
This thermal effect can be included through an additional equation 
that determines the rate of change of  the internal energy of the system.
\footnote{ In Ref.\cite{gps96}, strangeness production in quark
matter has been considered with allowance for the time dependence 
of the temperature. It has been shown that the temperature 
changes appreciably with time up to $\sim $50 MeV during the nonequilibrium
process, and that 
 the time scale for saturation of the strangeness in this case 
is smaller by orders of magnitude than the case 
where the temperature is held fixed. 
} 
A significant increase in the temperature could possibly lead to 
further reduction of the time scale for the growth of a condensate  
represented by the coherence time. 
On the other hand, as a result of the raise in the temperature, the critical density
$n_B^C$ for kaon condensation can be pushed up to higher density, 
which may cause the delay in the
onset of condensation. In this respect, it is necessary to obtain a phase diagram in
the $T-n_B$ plane\cite{ty98,yt99}, and the dynamics of the onset and growth of
condensation  must be considered along a trajectory in this plane. 

\section{Acknowledgements}
\label{sec:ack}

The authors wish to thank M. Yasuhira for discussions. 
We are grateful to Professor R.T.Deck for comments on the manuscript. 
One of the authors (T.M.) thanks for
support through the Grant-in-Aid of  Chiba Institute of Technology (C.I.T). Numerical
calculations  were carried out on the DEC Alpha Server 4100 System, C.I.T. 
This material is based upon work
supported in part by the  National Science Foundation through the Theoretical Physics Program 
under Grant Nos.PHY9008475 and PHY9722138, and by the Japanese 
Grant-in-Aid for Scientific Research Fund of the Ministry of Education, 
Science, Sports and Culture (08640369,11640272). 

\appendix
\section{Asymptotic behavior near chemical equilibrium}
\label{sec:appen}

We derive an expression for the relaxation time which characterizes 
the asymptotic behavior of the system near chemical equilibrium (stage III). 

\subsection{Linearization of $\xi^{({\rm KU})}(t)$ and $\xi^{({\rm MU})}(t)$}
\label{subsec:linear}

In stage III, both the forward and backward KT reaction rates are almost
equal, i.e., $\xi^{({\rm KT})}(t)\simeq 0$. Therefore, we consider the case where 
the relaxation proceeds only through  the KU and MU reactions, neglecting the KT
reaction.  The parameters $\xi^{({\rm KU})}(t)$ and 
$\xi^{({\rm MU})}(t)$ are written as 
\begin{mathletters}\label{axi}
\begin{eqnarray}
\xi^{({\rm KU})}(t)&=&\lbrack\delta\mu_e(t)-\delta\mu_K(t)\rbrack/T  \ ,
\label{axiku} \\
\xi^{({\rm MU})}(t)&=&\lbrack\delta\mu_e(t)
+\delta\mu_p(t)-\delta\mu_n(t)\rbrack/T \ ,
\label{aximu} 
\end{eqnarray}
\end{mathletters}
where $\delta\mu_i(t)$ ($i=p,n,e^-,K^-$) is the deviation of the chemical
 potential from the equilibrium value, i.e.,
$\delta\mu_i(t)=\mu_i(t)-\mu_i^{\rm eq}$. 
From the relation, $\mu_e=(3\pi^2n_e)^{1/3}$, one obtains 
\begin{equation}
\delta\mu_e(t)=(\pi/\mu_e^{\rm eq})^2\delta n_e(t) \ ,  
\label{delmue}
\end{equation}
where $\delta n_e(t)=n_e(t)-n_e^{\rm eq}$. 
One can also write the deviations of the chemical potentials $\delta\mu_K(t)$ and
$\delta\mu_p(t)-\delta\mu_n(t)$ in terms of $\delta n_i(t)$ 
(=$n_i(t)-n_i^{\rm eq}$) from linearized forms of 
the three equations with respect to $\delta\mu_K(t)$,
$\delta\theta(t)$, $\delta\mu_p(t)-\delta\mu_n(t)$, and
$\delta n_i(t)$ represented by: 
the expression for the kaon number density  [Eq. (\ref{nk})], the classical field
equation [Eq. (\ref{fieldeq}) ],  and the expression for the difference between
$\mu_p$ and $\mu_n$  [Eq. (\ref{ceppn})] .  Note that we can 
neglect the deviation of the thermal kaon part in the kaon number density,
i.e., $\delta n_K(t)=\delta\zeta_K(t)$ ($\delta n_K^T (t)=0$), because the
thermal kaon number density
$n_K^T(t)$  is almost constant after a condensate appears 
until the system reaches chemical equilibrium.  

The deviation of the  electron number density $\delta n_e(t)$ is 
divided into two contributions:
\begin{equation}
\delta n_e(t)=\delta^{({\rm KU})}n_e(t)+\delta^{({\rm MU})}n_e(t) \ , 
\label{delni}
\end{equation}
where $\delta^{({\rm KU})}n_e(t)$ [$\delta^{({\rm MU})}n_e(t)$] 
is caused by the KU reactions (the MU reactions). 
The number densities for the other species are related to $\delta^{({\rm
KU})}n_e(t)$ and $\delta^{({\rm MU})}n_e(t)$ by 
\begin{mathletters}\label{relation}
\begin{eqnarray}
\delta n_K(t)&=&-\delta^{({\rm KU})}n_e(t)  \ ,  \label{relation1} \\
\delta n_p(t)&=&\delta^{({\rm MU})}n_e(t) \ , \label{relation2} \\
\delta n_n(t)&=&-\delta n_p(t)=-\delta^{({\rm MU})}n_e(t)
\end{eqnarray}
\end{mathletters}

Following the above results, the deviation parameters (\ref{axi})
can be expressed in terms of $\delta^{({\rm KU})}n_e(t)$ and $\delta^{({\rm
MU})}n_e(t)$.  The result is 
\begin{mathletters}\label{axii}
\begin{eqnarray}
\xi^{({\rm KU})}(t)&=&\frac{p}{T}\delta^{({\rm KU})}n_e(t) 
+\frac{q}{T}\delta^{({\rm MU})}n_e(t) \ , \label{axii1} \\ 
\xi^{({\rm MU})}(t)&=&\frac{q}{T}\delta^{({\rm
KU})}n_e(t)  +\frac{r}{T}\delta^{({\rm MU})}n_e(t) \ ,
\label{axii2}  
\end{eqnarray}
\end{mathletters}
where
\begin{mathletters}\label{coe}
\begin{eqnarray}
p&=&\Bigg\lbrack\Big(\frac{\pi}{\mu_e}\Big)^2+
\frac{1}{f^2\Big\{\sin^2\theta+4(\cos\theta+b/\mu_K)^2\Big\}}\Bigg\rbrack^{\rm
eq} \ , \label{coe1} \\
q&=&\Bigg\lbrack\Big(\frac{\pi}{\mu_e}\Big)^2+
\frac{1+\cos\theta+2b/\mu_K}{2f^2\Big\{\sin^2\theta+
4(\cos\theta+b/\mu_K)^2\Big\}}\Bigg\rbrack^{\rm
eq} \ , \label{coe2} \\
r&=&\Bigg\lbrack\Big(\frac{\pi}{\mu_e}\Big)^2
+\frac{(1-\cos\theta)(\cos\theta+2b/\mu_K)}{2f^2\Big\{\sin^2\theta+
4(\cos\theta+b/\mu_K)^2\Big\}}
+\frac{2}{3}\frac{(3\pi^2)^{2/3}}{2m_N}(n_p^{-1/3}+n_n^{-1/3})
+\frac{8V_{\rm sym}(n_{\rm B})}{n_{\rm B}} \Bigg\rbrack^{\rm eq} 
\ . \label{coe3}
\end{eqnarray}
\end{mathletters}
In Eq.(\ref{coe}), the first term comes from the deviation of the  
electron chemical potential 
$\delta\mu_e$, (\ref{delmue}), and the second term from the deviation of the
parameters  for a condensate, i.e., $\delta\mu_K$ and $\delta\theta$. 
The third and fourth terms in
Eq. (\ref{coe3}) come from the deviation of the proton and neutron chemical
potentials.  
It is to be noted that a condensate already develops fully in stage III, 
so that the quantities for the condensate are almost constant. 
Thus it is a good approximation to set $\delta\mu_K=\delta\theta=0$, 
which leads to $\displaystyle p=q\simeq 
\Big(\frac{\pi}{\mu_e^{\rm eq}}\Big)^2$,  and $\displaystyle r\simeq
\Bigg\lbrack\Big(\frac{\pi}{\mu_e}\Big)^2 +\frac{2}{3}\frac{(3\pi^2)^{2/3}}{2m_N}(n_p^{-1/3}+n_n^{-1/3})
+\frac{8V_{\rm sym}(n_{\rm B})}{n_{\rm B}} \Bigg\rbrack^{\rm eq}$. 

\subsection{Differential equations for $\xi^{({\rm KU})}(t)$ and 
$\xi^{({\rm MU})}(t)$}

By differentiating both sides of Eq. (\ref{axii}) with time, one finds that 
the rates of change of $\xi^{(\rm KU)}(t)$ and $\xi^{({\rm MU})}(t)$ are related
to the reaction rates for KU and MU, through the rate equations 
\begin{mathletters}\label{arate}
\begin{eqnarray}
\frac{d}{dt}\delta^{({\rm KU})}n_e(t) 
&=&\Gamma^{({\rm KU-F})}(\xi^{({\rm KU})}(t),T)
-\Gamma^{({\rm KU-F})}(-\xi^{({\rm KU})}(t),T) \ , \label{arate1} \\
\frac{d}{dt}\delta^{({\rm MU})}n_e(t) 
&=&\Gamma^{({\rm MU-F})}(\xi^{({\rm MU})}(t),T)-\Gamma^{({\rm
MU-F})}(-\xi^{({\rm MU})}(t),T)  \label{arate2}  \ . 
\end{eqnarray}
\end{mathletters}
For the system close to chemical equilibrium, $|\xi^{({\rm KU})}(t)|\ll 1$ and 
$|\xi^{({\rm MU})}(t)|\ll 1$. In this case, the r.h.s. of Eqs. (\ref{arate1}) and 
(\ref{arate2}) can be
expanded up to first order in $\xi^{({\rm KU})}(t)$ 
and $\xi^{({\rm MU})}(t)$, respectively. Thus one obtains 
the coupled differential equations for $\xi^{({\rm KU})}(t)$ and 
$\xi^{({\rm MU})}(t)$: 
\begin{mathletters}\label{diff}
\begin{eqnarray}
\frac{d}{dt}\xi^{({\rm KU})}(t)&=&-\frac{p}{T}
\widetilde\Gamma^{({\rm KU-F})}\cdot\xi^{({\rm KU})}(t)
-\frac{q}{T}\widetilde\Gamma^{({\rm MU-F})}\cdot\xi^{({\rm MU})}(t) \ ,
\label{diff1} \\
\frac{d}{dt}\xi^{({\rm MU})}(t)&=&-\frac{q}{T}
\widetilde\Gamma^{({\rm KU-F})}\cdot\xi^{({\rm KU})}(t)
-\frac{r}{T}\widetilde\Gamma^{({\rm
MU-F})}\cdot\xi^{({\rm MU})}(t) \ , \label{diff2}
\end{eqnarray}
\end{mathletters}
where
\begin{mathletters}\label{tildegamma}
\begin{eqnarray}
\widetilde\Gamma^{({\rm KU-F})}&\equiv& 2\frac{|I_2'(0)|}{I_2(0)}
\Gamma^{({\rm KU-F})}(\xi^{({\rm KU})}=0, T)
\simeq 1.3 \times \Gamma^{({\rm KU-F})}(\xi^{({\rm KU})}=0, T)\propto T^5 \ ,
\label{tildegamma1} \\
\widetilde\Gamma^{({\rm MU-F})}&\equiv &
2\frac{|J_2'(0)|}{J_2(0)}\Gamma^{({\rm MU-F})}(\xi^{({\rm MU})}=0, T)
\simeq 1.2 \times \Gamma^{({\rm MU-F})}(\xi^{({\rm MU})}=0, T)\propto T^7 \ , 
\label{tildegamma2} 
\end{eqnarray}
\end{mathletters}
with
$ I_2(0)=\frac{3}{2}\pi^2\zeta(3)+\frac{45}{2}\zeta(5)$=41.13,
$ I_2'(0)\equiv (dI_2(u)/du)_{u=0}=-\frac{17}{60}\pi^4$, 
$ J_2(0)=\frac{27}{2}\pi^4\zeta(3)
+225\pi^2\zeta(5)+\frac{2835}{4}\zeta(7)$=4598,
and
$ J_2'(0)\equiv (dJ_2(u)/du)_{u=0}=-\frac{367}{126}\pi^6$.  
From Eq. (\ref{diff}), one obtains the following forms of the solutions:
\begin{mathletters}\label{asol}
\begin{eqnarray}
\xi^{({\rm KU})}(t)&=&C_1e^{-t/\tau_a}+C_2 e^{-t/\tau_b} \ , \label{asol1} \\
\xi^{({\rm MU})}(t)&=&C_3 e^{-t/\tau_a}+C_4 e^{-t/\tau_b} \ , \label{asol2}
\end{eqnarray}
\end{mathletters}
where $C_1-C_4$ are constants, and $\tau_a$ and $\tau_b$ are time constants
which characterize the asymptotic evolution of the system.  
The relaxation time $\tau_{\rm rel}$ is defined as the larger of $\tau_a$ and
$\tau_b$, and one obtains
\begin{equation}
{\footnotesize
\tau_{\rm rel}=\frac{T}{2}\frac{p\widetilde\Gamma^{({\rm
KU-F})}+r\widetilde\Gamma^{({\rm MU-F})}
+\Big\lbrack\{p\widetilde\Gamma^{({\rm
KU-F})}-r\widetilde\Gamma^{({\rm MU-F})}\}^2
+4q^2\widetilde\Gamma^{({\rm KU-F})}\widetilde\Gamma^{({\rm
MU-F})}\Big\rbrack^{1/2} }{(pr-q^2)\widetilde\Gamma^{({\rm KU-F})}
\widetilde\Gamma^{({\rm MU-F})}} \ .}
\label{atau}
\end{equation}

\begin{table}
\caption{\footnotesize Quantities for the initial noncondensed
state ($t=0$) and for the  chemically equilibrated  state ($t\rightarrow
\infty$). The former (the latter) quantities are denoted 
by the superscript `0' (`eq'). All the values are estimated at $T$=0. }
\label{tabcond} 
\begin{tabular}{c || c c | c c c}
$n_{\rm B}$({\rm fm}$^{-3}$) &$\mu_K^0 $ ( MeV ) & $x_p^0$  & 
$\theta^{\rm eq}$ (${\rm rad}$) & $\mu_K^{\rm eq} $ ( MeV )
& $x_p^{\rm eq}$ \\
\hline\hline 
 0.55 & $-139$ & 0.14 & 0.48 & 203  & 0.23 \\
\hline 
0.70 &  $-180$ & 0.16 & 0.91 & 114 & 0.39 \\ 
\end{tabular}
\end{table}~
\begin{table}
 \caption{\footnotesize The nucleation time $\tau_{\rm nucl}$, the coherence
time $\tau_{\rm coh}$, and the relaxation time $\tau_{\rm rel}$ for different
densities and temperatures. }
\label{tabtau} 
\begin{tabular}{ccccc}
$n_{\rm B} ({\rm fm}^{-3} )$ &T ({\rm K}) & $\tau_{\rm nucl}$ (${\rm sec}$) & 
$\tau_{\rm coh}$ (${\rm sec}$) & $\tau_{\rm rel}$ (${\rm sec}$) \\
\hline\hline 
0.55 & 1$\times$10$^{10}$ & 2$\times$10$^{-8}$ & 2$\times$10$^{-1}$ &
4$\times$10 \\
        & 1$\times$10$^{11}$ & 8$\times$10$^{-8}$ & 5$\times$10$^{-5}$ &
2$\times$ 10$^{-4}$ \\
        & 5$\times$10$^{11}$ & 6$\times$10$^{-9}$ & 4$\times$10$^{-8}$ &
5$\times$ 10$^{-7}$\\
\hline 
0.70 & 1$\times$10$^{10}$ & 1$\times$10$^{-11}$ & 6$\times$10$^{-2}$
& 9$\times$10 \\
        & 1$\times$10$^{11}$ & 3$\times$10$^{-10}$ & 1$\times$10$^{-5}$&
2$\times$ 10$^{-4}$ \\
        & 5$\times$10$^{11}$ & 3$\times$10$^{-10}$ & 2$\times$10$^{-8}$
& 1$\times$ 10$^{-7}$ \\ 
\end{tabular}
\end{table}

\end{document}